\documentclass[prb,twocolumn,showpacs,amsmath,amssymb,superscriptaddress]{revtex4}
\usepackage{graphicx}
\usepackage{graphics}
\usepackage{dcolumn}
\usepackage{bm}
\usepackage{makeidx}

\def\f{_{\mathrm{F}}}
\def\k{_{\mathrm{K}}}
\def\xx{_{\mathrm{xx}}}
\def\xy{_{\mathrm{xy}}}
                                                                                                                           
\begin{document}
\title{MAGNETO-TRANSPORT    AND    MAGNETO-OPTICAL    PROPERTIES    OF
FERROMAGNETIC (III,Mn)V SEMICONDUCTORS: A REVIEW}
\author{Jairo Sinova}
\affiliation{Texas A\&M University,  Physics Department,  4242 TAMU,  College Station, TX 77943-4242}
\author{T. Jungwirth}
\affiliation{University  of  Texas at  Austin,  Physics Department,  1
University   Station  C1600,   Austin  TX   78712-0264}
\affiliation{ Institute of Physics  ASCR, Cukrovarnick\'a 10, 162 53
Praha 6, Czech Republic }
\author{John \v Cerne}
\affiliation{Depatment of Physics, University at Buffalo, SUNY, Buffalo, New York 14260}
\date{\today}

\begin{abstract}
Rapid  developments in  material research  of  metallic ferromagnetic
(III,Mn)V semiconductors over  the past few years have  brought a much
better understanding  of these complex materials. We  review here some
of  the  main  developments  and  current understanding  of  the  bulk
properties  of these  systems within  the metallic  regime, focusing
principally  on the  magneto-transport and  magneto-optical properties.   
Although  several theoretical  approaches  are
reviewed,  the  bulk of  the  review  uses  the effective  Hamiltonian
approach, which has proven useful in describing many of
these  properties namely in (Ga,Mn)As  and (In,Mn)As. The model assumes
a ferromagnetic coupling between Mn d-shell local moments mediated
by holes in the semiconductor valence band.
\end{abstract}

\maketitle 
\tableofcontents

\section{Introduction}
The  artificial marriage  of ferromagnetic and  semiconducting  properties
is one of the main motivations behind the creation and study of diluted magnetic semicondcutors (DMS) focused on
(III,Mn)V  materials. This  effort is  being
fueled,  in  large  part,  by  the endless  demand  for  technological
advances in our society whose global interdependence in communications
and technology  is evident  throughout our daily  activities.  Moore's
law, predicting an exponential growth of computational ability, is now
believed to be  in perryl and there is a need  for a new technological
revolution if the fast pace of technology is to be mantained.

Such a technological  revolution will take place if  three major goals
can  be  achieved  in   the  development  of  ferromagnetic  (III,Mn)V
materials:  i)  room  temperature  ferromagnetism, ii)  dependence  of
transport and optical properties  on the magnetic state,
and iii)  that the materials mantain  their fundamental semiconducting
characteristics, i.e.  strong sensitivity to  doping, external fields,
and  light, and are compatible and integrable with conventional (non-magnetic)
micro- and opto-electronic semiconductor technologies.  
Since  the  first observation  of  ferromagnetism at  low
temperatures in  the (III,Mn)V  semicondcutors in the  early nineties,
\cite{Ohno:1992_}  steady progress  has been  taking place  in  each of
these  major  fronts. More  importantly,  over  the  last five  years,
jumpstarted  by  the discovery  of  ferromagnetism  in (Ga,Mn)As  with
ferromagnetic temperatures in excess  of 100 K, \cite{Ohno:1998_a} the
efforts  in   this  field  have  increased   dramatically  in  several
coordinated  fronts,  both experimental  and  theoretical. This  rapid
research has yielded many advances,  prompting the need for an overall
review  of at  least several  fast evolving  parts of  this scientific
effort. In this  article we discuss   transport and magento-optical
properties of these  materials focusing
on the  metallic regime and on theoretical approaches based
on the
effective Hamiltonian model which  have been succesful in obtaining
qualitative and  quantitative understanding of various transport 
characteristics of these systems. Other
important areas, such as  the insulating low-doping regime have been
covered                            by                            other
reviews.\cite{DasSarma:2003_cond-mat/0304219,Timm:2003_}
Furthermore, we will  narrow our focus on the  most widely studied and
well  understood materials, (Ga,Mn)As  and (In,Mn)As,  only mentioning
briefly some  of the current prevailing  ideas in regards  to the much
lesser understood ones such as the Mn-doped nitrides or phosphides.

As  in any  review,  a preference must  be  given to  the more  relevant
publications  containing the  most  recent developments  and the  ones
which  have  had the  largest  impact  in  understanding these  complex
materials.  Unless the  reference section  is   equal in  length to the
actual review,  it is  impossible to cite  every contribution  to this
field. To  remedy 
omissions that may have partly originated from the authors' own work
perspective, we refer to  a  more extended  and precise
database   of   published  and   pre-print   works maintained
within the   "Ferromagnetic
Semiconductors  Web Project" at  http://unix12.fzu.cz/ms.  
This  web  page  contains  a large  body  of  theoretical  and
experimental  results (that can  be obtained  in several  formats), as
well  as the  most extended  bibliography database  linked  to diluted
magnetic (III,V) semicondcutors that we are aware of. We encourage the
reader in need of a more detailed bibliography to use this resource.
  
We  also note  that a  few  review articles  on other  aspects of  the
physics of DMS, which we will not discuss extensivelly in this review,
have appeared  over the  past few  years and may  help the  reader who
wants to  extend the narrow scope  of this article.  Given however the
rapid developments over the past two years alone, the reader should be
warned that these  reviews, although excelent, may not  sufice to give
the   most   current   experimental   developments   and   theoretical
understanding  and  justifications.  Several  extended papers  on  the
physics of  ferromagnetism and proposed  mechanisms, some of  which we
briefly         review        here,        are         shown        in
Refs. ~\onlinecite{Ohno:1998_a,Ohno:1999_a,Dietl:2002_b,Matsukura:2002_}. 
A theoretical
review  of the widely  used phenomenological  Zener mean  field model,
highly succesful  in (Ga,Mn)As and (In,Mn)As studies,  is presented in
Ref.  ~\onlinecite{Konig:2003_} and ~\onlinecite{Dietl:2001_b}.  
A large
body  of  work up  to  2001  in  density functional  first  principles
calculations    in    (III,V)     DMS    is    also    available    in
Ref.  ~\onlinecite{Sanvito:2002_b}. Another theoretical  approach based
on lattice models and dynamical mean field treatment of disorder, most
succesful    in    the    low-doping    regime,   is    reviewed    in
Ref.  ~\onlinecite{DasSarma:2003_cond-mat/0304219}. The  ever important
effects of  disorder, omnipresent in  these materials, is  reviewed in
Ref.  ~\onlinecite{Timm:2003_}.  A  general  discussion
behind  the technological  motivation for  DMS has  been  presented in
~\onlinecite{Wolf:2001_},                     ~\onlinecite{Dietl:2001_a},
~\onlinecite{Gregg:2002_}, and  ~\onlinecite{Ohno:2002_b}, where optical
isulators and magnetoresistance elements are explored.
                                                                                                                          
We organize the  rest of this article as  follows: In Sec. \ref{basic}
we present a short general overview of DMS from the experimental facts
point of view. In Sec. \ref{theory} we briefly discuss the theoretical
model  used  and  its  experimental  basis to  describe  the  metallic
regime. Sec.  \ref{transport} reviews the transport  properties of DMS
such  as  diagonal  resisticity,  magneto-resistance  anisotropy,  and
anomalous Hall effect. Sec.  \ref{optics} covers the different optical
and  magneto-optical  properties  of   the  systems  such  as  optical
absorption,   magnetic-circular  diachroism,  Raman   scattering,  and
cyclotron resonance. A short summary and outlook with different open questions 
is given in section V.

\subsection{The basic picture of DMS}
\label{basic}
The basic idea behind  creating these novel ferromagnetic materials is
simple: based on the lower valence  of Mn, substituing Mn in a (III,V)
semiconductor for the cations can dope the system with holes; beyond a
concentration of   1\% there are enough induced  holes to mediate a
ferromagnetic coupling between  the local $S=5/2$ magnetic moments
provided  by the  strongly  localized  3d$^5$ electrons  of  Mn and  a
ferromagnetic ordered state can  ensue. This rudamentary but generally
correct picture  of ferromagnetism in  many (III,Mn)V materials  is of
course aquired by peacing  together the different experimental results
obtained   through  different   characterization  techniques   of  the
materials. The simplicity of the model hides within it a cornucopia of
physical effects  present in  these materials such  as metal-insulator
transitions,   carrier  mediated  ferromagnetism,   disorder  physics,
magneto-resistance effects, magneto-optical effects, intricate coupled
magnetization dynamics, post-growth dependent properties, etc.

In any doped semiconductor, an understanding of its properties must be
preceeded by a  study of the doping impurities.   Since, in this case,
the Mn impurity dopants provide the local magnetic moments responsible
for ferromagetism  and, at  the same time,  provide the  hole carriers
which mediate the exchange coupling between the impurities themselves,
understanding the  nature of the  impurities and the  different states
that  they induce is  even more  fundamental in  order to  establish a
relevant theoretical model.

\subsection{Impurities in DMS}

Under equilibrioum  growth conditions, the  solubility of Mn  in III-V
semicondcutor  crystals has  an upper  limit of  0.1 \%.   Beyond this
concentration phase separation and surface segregation takes place. To
circumvent  this  problem,   low-temperature  molecular  beam  epitaxy
(LT-MBE) technique  was applied leading to the  first succesful growth
of InAs and GaAs based DMS.\cite{Ohno:1992_}  The most common and stable
position of Mn in the host semiconductor lattice is on
the Ga site.~\cite{Battacharjee:2000_,Linnarsson:1997_,Schneider:1987_,Erwin:2002_,Masek:2002_cond-mat/0201131}

An important initial question is what is the neutral state
of the Ga-substitutional
Mn impurity.    Experimentally,   through  electron   paramagnetic
resonance                (EPR)               and               optical
measurements,\cite{Linnarsson:1997_,Szczytko:2001_,Szczytko:1999_b}
for even  very few percentage  of Mn (lower  by an order  of magnitude
than  the concentration  marking the  matal-insulator  transition) the
only impurity  level observed  is the d5-Mn$^{2+}$  ionized state,
indicating  that indeed  the  strong localized  3d$^5$ electrons  with
total  spin  $S=5/2$  are   a  good  starting  point  for  theoretical
models. The corresponding 3d$^5$ plus a weakly bound hole neutral state
is observed experimentally only within a narrow range of Mn concentrations
due to a complete compensation in the low Mn doping regime and due to
the Mott
insulator-to-metal transition at high doping.
Other  candidates for the  Mn neutral impurity such  as 3d$^4$
states,  important for  double-exchange models  to be  applicable, are
    not     observed     experimentally.\cite{Linnarsson:1997_}
Furthermore,  the  itinerant holes originating from
the Mn acceptor level  have  also  been
observed           through          photoemission          experiments
\cite{Okabayashi:1999_,Okabayashi:2001_}  to  be  associated with  the
ones of the host semiconductor valence band, i.e. they have As 4p-character,
which provides  further evidence for the shallow acceptor nature of
the substitutional Mn impurity in GaAs and InAs.

The  other impurities  present because  of the  non-equilibrium growth
process,  As-antisites  and  Mn-interstitials, compensate  the  induced
holes and  therefore  reduce  the free carrier  concentration with
respect  to the  substitutional-Mn density.  Annealing  procedures, at
temeratures  slightly  lower than  growth  temperature,  have shown  a
reduction            of            the           ammount            of
compensation.\cite{Ohno:1999_a,Edmonds:2002_b,Ku:2003_,Mathieu:2002_cond-mat/0208411,Yu:2002_a,Edmonds:2004_,Kuryliszyn-Kudelska:2003_cond-mat/0304622}
The  initial procedure  \cite{Ohno:1999_a}  has now  been modified  by
different                                                        groups
\cite{Edmonds:2002_b,Ku:2003_,Mathieu:2002_cond-mat/0208411,Kuryliszyn-Kudelska:2003_cond-mat/0304622}
and the carrier concentration can actually be tuned precisedly through
resistance-monitored  annealing.  \cite{Edmonds:2002_b} Two possible  
mechanism
for the  reduction of compensation  have been considered:  either the
number of As-antisites is reduced  through migration of  As atoms to
their  correct sites,  or the  Mn-interstitials 
migrate  to  the  surface  or  to the  substitutional  sites.  Early
theoretical models assumed the  former scenario, however, a recent key
Rutherford backscattering experiment\cite{Ku:2003_}  has shown that it
is the reduction of Mn-interstitials and their out-diffusion
to the surface \cite{Edmonds:2004_} which is responsible for lowering
the ammount of compensation.

\subsection{Other experimental observations in DMS}

Besides the  experimental results revieling  what is happening  in the
complicated  annealing process and  the nature  of the  impurities, there
are other experimental observations that provide key clues for understanting
the origins and fundamental properties of (Ga,Mn)As and (In,Mn)As DMS:

a) The ferromagnetic behavior of  DMS materials is only observed above
a  critical doping  level of  about 1\%.  \cite{Ohno:1999_a}  At lower
doping the compensation is  almost complete and the mediating carriers
needed to exchange-couple the localized Mn moments are not present and
therefore no ferromagnetic ordering occurs.

b) The  carrier-induced ferromagnetic nature of the  ordered state has
been    demonstrated    by    field-effect    (In,Mn)As    experiments
\cite{Ohno:2000_a} where the carrier concentration was tuned by a gate
and the corresponding critical temperature was modified accordingly.

c) Strain effects, due to the lattice mismatch between the DMS
layer and the substrate,
influence    the   magnetic    anisotropy    of   the    ferromagnetic
state. It has been shown that the ferromagnetic easy axis can be 
along  the growth direction or in the plane 
depending on  whether the strain is tensile  or compressive.
\cite{Ohno:1999_a,Dietl:2001_b} Such  phenomena  are interpreted
in terms of well  known strain  effects in  the
spin-orbit coupled valence bands of the host semiconductor.
 
c)  Both magneto-optical  effects, e.g.  magnetic  circular diachroism
(MCD),\cite{Szczytko:1999_a}  and resistivity  measurements  above the
critical temperature,\cite{Omiya:2000_} indicate an anti-ferromagnetic
coupling  between the  local  3d$^5$  electrons and  the
valence hole. This so-called kinetic-exchange interactions originates
from the hybridization between the Mn 3d-orbitals and the neighboring
As 4p-orbitals and is much larger than the pure Coulomb exchange 
interaction.\cite{Dietl:1994_}

e)  In transport measurements, 
a  large     anomalous-Hall   effect  (AHE) (see
Sec. \ref{AHE})  completely dominates the low-field
off-diagonal resistance coefficient. The strong
intrinsic spin-orbit coupling present  in the host semiconductor
valence band can account for the measured magnitude and sign of the
AHE.

There  are of  course many  other relevant  experiments which  are not
higlighted  above  which  must  be  taken into  account.  However  the
interpretation of some or most  of them are still being debated, since
they seem to depend critically  on the annealing procedures and can be
interpreted in  several ways. The  above results are considered  to be
general and well established and  will likely not change in  the newly  grown
samples as the materials development progresses.

\section{Theoretical models of DMS}
\label{theory}

The  modeling of  a collective behavior  of interacting  electrons  is a
complicated task and, in many instances, must be guided by experimental
evidence of  the low energy  degrees of freedom  in order to  obtain a
correct minimal  model which will  capture the observed  effects and
will make useful predictions. With
this  in   mind,  there  are  typically   three  approaches,  somewhat
complementary,  used   to  describe  DMS   systems:  first  principles
density-functional  theory (DFT)  and microscopic
tight-binding  models, effective
Hamiltonian models, and lattice models.

DFT  is   an  important  tool  for  studying   microscopic  origins  of
ferromagnetism  through  calculations  of  electronic,  magnetic,  and
structural     ground-state     properties.\cite{Sanvito:2002_b}     A
local-density-approximation   (LDA)   of   the  DFT,   combined   with
disorder-averaging coherent-potential approximation (CPA) or supercell
approach, has been used successfully to address physical parameters of
(III,Mn)V DMS that are derived from total-energy calculations, such as
the  lattice constants,\cite{Masek:2003_}  and  formation and  binding
energies                           of                          various
defects.\cite{Masek:2002_cond-mat/0201131,Erwin:2002_}   However,   in
Mn-doped  DMS,  LDA fails  to  account  for  strong correlations  that
suppress fluctuations in the number  of electrons in the d-shell. As a
result, the  energy splitting between the occupied  and empty d-states
is  underestimated which  leads  to an  unrealistically large  d-state
local DOS near  the top of the valence band and  to an overestimate of
the             strength            of             the            sp-d
hybridization.\cite{Akai:1998_,Sanvito:2002_b}  This shortcomings have
been corrected  recently by  LDA+U and self-interaction  corrected LDA
schemes,  which account for  correlations among  Mn 3d  electrons, and
have  been used  to  obtain  more realistic  energy  spectra and  show
agreement with the experimental observation that the valence holes
have mostly As 4p-character.\cite{Park:2000_}

A practical approach, that circumvents some of the complexities of this
strongly-correlated many-body problem is the microscopic
tight-binding (TB) band-structure
theory. Within the model,
local changes of the crystal potential at Mn and other impurities
are represented by shifted atomic levels. A proper parameterization of these
shifts, of the Hubbard correlation potential that favors single
occupancy of the localized d-orbitals, the Hund potential forcing
the five d-orbital spins to align,  and of the hopping amplitudes between 
neighboring atoms
provides correct band gap for the
host III-V semiconductor and an appropriate exchange splitting of the Mn
d-levels. The TB model is a semi-phenomenological theory,
however, it shares the virtue with first principles
approaches of treating disorder
microscopically. The decoherence of Bloch quasiparticle states or effects of
doping and
disorder on the strength of the sp-d exchange coupling and effective
Mn-Mn interaction are among the problems
that have been analyzed using this tool.
\cite{Masek:2002_b,Blinowski:2003_,Jungwirth:2003_a}

In  the  metallic regime,  where  the  largest critical  ferromagnetic
temperatures  are   achieved  (  for   doping  levels  above   1.5\%
\cite{Campion:2003_b}),
semi-phenomenological models  that are built  on Bloch states  for the
band  quasiparticles, rather than  localized basis  states appropriate
for  the  localized   regime,\cite{Bhatt:2002_}  provide  the  natural
starting point  for a model  Hamiltonian which reproduces many  of the
observed  experimental effects.   Recognizing that  the  length scales
associated with  holes in the DMS  compounds are still  long enough, a
${\bf  k}   \cdot  {\bf  p}$ envelope  function   description  of  the
semiconductor   valence  bands   is  appropriate.    Since   for  many
properties, e.g. anomalous Hall  effect and magnetic anisotropy, it is
necessary to incorporate intrinsic  spin-orbit coupling in a realistic
way, the six-band (or multiple-band, in general)
Kohn-Luttinger  (KL) ${\bf k} \cdot  {\bf p}$
Hamiltonian   that  includes   the  spin-orbit   split-off   bands is
desirable.  \cite{Abolfath:2001_b,Dietl:2001_b}  The approximation  of
using the KL Hamiltonian to describe the free holes is based primarely
in  the  shallow  acceptor  picture demonstrated  by  the  experiments
\cite{Battacharjee:2000_,Linnarsson:1997_,Schneider:1987_}           in
(Ga,Mn)As  and (In,Mn)As  and must  be re-examined  for any  other DMS
materials that this model is applied to.

Besides the KL Hamiltonian parameters of the host III-V compound which
have long  been established,\cite{Vurgaftman:2001_}
the phenomenological part  of the strategy
follows from asserting, rather  than deriving, the localized nature of
the   Mn  d-orbital   moments   and  from   parameterizing  the   sp-d
hybridization  by   an  effective  exchange   constant  $J_{pd}$.  The
localization  assumption  is  again  verified  by  electron  resonance
experiments \cite{Battacharjee:2000_,Linnarsson:1997_,Schneider:1987_}
and the value of $J_{pd}$ is obtained from resistivity measurements in
the     paramagnetic     regime     \cite{Omiya:2000_}     and     MCD
measurements. \cite{Szczytko:1999_a}  Hence, the effective Hamiltonian
considered within this model is
\begin{equation}
{\cal H}={\cal H}_{KL}+J_{pd}\sum_{I} {\bf S}_I\cdot \hat{\bf s}({\bf r}) \delta({\bf r}-{\bf R_I})+{\cal H}_{dis},
\label{Heff}
\end{equation}
where ${\cal  H}_{KL}$ is the  six-band (multiple-band) Kohn-Luttinger (KL)
${\bf k} \cdot  {\bf p}$ Hamiltonian, \cite{Konig:2003_,Dietl:2001_b}
the second  term is  the short-range antiferromagnetic  
kinetic-exchange interaction
between local spin ${\bf S}_I$  at site ${\bf  R}_I$ and the itinerant
hole spin (a finite
range  can  be incorporated  in  more  realistic  models), and  ${\cal
H}_{dis}$  is   the  scalar  scattering   potential  representing  the
difference  between a  valence  band electron  on  a host  site and  a
valence  band  electron  on  a   Mn  site  and  the  screened  Coulomb
interaction of the itinerant electrons with the ionized impurities.

Several approximations can  be used to vastly simplify the  above model, 
namely,
the virtual crystal approximation  (replacing the spatial dependence of
the  local  Mn   moments  by  a  constant  average)   and  the mean  field
theory description.\cite{Konig:2003_,Dietl:2001_b}  
In the  metallic
regime, the disorder can be treated by a Born approximation or by more
sophisticated,        exact-diagonalization       or       Monte-Carlo
methods.~\cite{Jungwirth:2002_c,Sinova:2002_,Yang:2003_,Schliemann:2002_b,Schliemann:2002_a}
The  effective Hamiltonian in Eq. \ref{Heff} 
allows  to  use standard  electron-gas  theory tools  to
account  for hole-hole  Coulomb interactions.  This  envelope function
approximation  model is also suitable for studying magnetic  semiconductor
heterostructures,       like       superlattices, quantum
wells and digitally doped layers.\cite{Lee:2002_a}  
The  validity  of such  semi-phenomenological
Hamiltonian,  which does  not  contain any  free  parameters, must  be
confirmed ultimately  by experiments.   Its accurate
description  of   many  thermodynamic  and   transport  properties  of
metallic (Ga,Mn)As samples, such as the
measured   transition  temperature,\cite{Dietl:2000_,Jungwirth:2002_b}
the anomalous  Hall effect,\cite{Jungwirth:2003_b,Jungwirth:2002_a} the
anisotropic magnetoresistance,\cite{Jungwirth:2003_b,Jungwirth:2002_c},
the magneto-crystalline anisotropy,\cite{Abolfath:2001_b,Konig:2001_b}
the spin-stiffness \cite{Konig:2001_b},  the ferromagnetic domain wall
widths, \cite{Dietl:2001_c} the magnetic dynamic damping coefficients,
\cite{Sinova:2003_cond-mat/0308386}     and     the    magneto-optical
properties,   \cite{Dietl:2001_b,Sinova:2002_,Sinova:2003_}  has proven
the merit of this effective Hamiltonian approach. 

One has  to keep in  mind, however, that as  any semi-phenomenological
model it  may fail to  capture the correct  physics that leads  to the
ferromagnetic  phase  in some  materials  or  in  a certain  range  of
parameter  values.  Such  models  can  only  be  verified  by  careful
comparison with  experiments and tested through  their predictions and
agreement  with experimetnts.   For example,  Mn-doped  nitride and
phosphide compounds  or insulating DMS samples  with low concentration
of  Mn ions  require a  theoretical description  that goes  beyond the
picture  of the host  band quasiparticles  that are  weakly hybridized
with  the  localized  Mn  d-electrons.   Particularly nitrates  are  not
believed to be well  modeled by this semi-phenomenological Hamiltonian
since  Mn is a deep acceptors in this case and and charge fluctuations
on the Mn d-levels may play important role.

There has also been theoretical  work on (III,Mn)V DMS materials based
on a still  simpler model where holes are assumed  to hop only between
Mn  acceptor  sites, where  they  interact  with  the Mn  moments  via
phenomenological                                               exchange
interactions.\cite{Chattopadhyay:2001_,DasSarma:2003_cond-mat/0304219}
These  models have  the advantage  of approaching  the physics  of the
insulating dilute  Mn limit,  and can also  be adapted to  include the
holes  that are  localized on  other  ionized defects  besides the  Mn
acceptors through dynamical mean field (or CPA) techniques.  
The free-parameter
nature  of  this phenomenological  approach  and their  oversimplified
eletronic  structure  allows  to  make only  qualitative  predictions,
however, and the models are also not appropriate for studying the high
$T_c$ metallic samples.

\section{Transport properties of DMS systems}
\label{transport}
The  different  transport  coefficients  of DMS  and  their  magnetic,
temperature, and  material composition  dependance have been  the most
important and  widely used characterization tools of  DMS. Besides the
diagonal resistivity  which indicates metal-insulator  transitions and
possible   critical   behavior   at   the   ferromagnetic   transition
temperature, other  material transport properties  such as anisotropic
magnetoresistance,  anomalous  and  ordinary  Hall effect,  and  giant
magnetoresistance  have been used  to both  characterize and  test the
different  theoretical models of  DMS materials.   In this  section we
consider the diagonal conductivity  general features, how the exchange
coupling between the localized  moments and free carriers is extracted
within the  paramagnetic regime, the doping  and carrier concentration
conductivity   dependence  at   low   temperatures,  the   anisotropic
magnetoresistance, and the anomalous
Hall  effect.   Throughout,  we  will  focus  on   the  comparison  of
theoretical  models  (mainly  the  semi-phenomenological  effective
Hamiltonian model  relevant to the  metallic regime) to  the different
experimental observations.

\subsection{General features of resistivity in DMS}

DMS materials  exhibit an insulating or metallic  behavior (defined by
the resistivity  in the  limit of zero  temperature) depending  on its
doping  level  and  post-growth  annealing  procedures.  
In  as-grown  samples,
metallic behavior  is typically observed for  a  range  of 2-5\% Mn
doping and  an insulating  behavior for higher  and lower  doping than
this range.\cite{Matsukura:1998_} In  addition to this metal-insulator
quantum transition,  the resistivity  as  a  function of  temperature
typically  exhibits   a  peak  or  shoulder   near  the  ferromagnetic
transition  temperature  for   both  insulating  (peak)  and metallic
(shoulder)
samples.~\cite{Potashnik:2002_cond-mat/0212260,VanEsch:1997_,Matsukura:1998_,Hayashi:1997_}
The  non-monotonic   behavior  near   $T_c$  is
typically associated with critical-scattering but so far no theory has
been developed which explains such behavior in a qualitative or
 quantitative way.  Typically the Fisher-Langer theory 
of correlated fluctuations is invoked, however, it
predicts an infinite  derivative of the resistivity at  $T_c$, which is
clearly not the case in any studied (Ga,Mn)As DMS sample.
There exists also a drastic  reduction of  the resistivity upon  annealing, associated
with the increase of the  carrier concentration and to a lesser extent
a   reduction    of   the    disorder   scatterers.~\cite{Edmonds:2002_b,Potashnik:2002_cond-mat/0212260,VanEsch:1997_,Matsukura:1998_,Hayashi:1997_} 
The on-set  of  the  metal-insulator  transition  at 1.5\% Mn doping
is  close  to  the  Mott
insulator  limit of  a doped  semiconductor  similar to  Si:P 
and the optimally annealed samples remain metallic throughout the
whole range of Mn concentrations above 1.5\%.\cite{Campion:2003_b}

\begin{figure}
\includegraphics[width=3.4in]{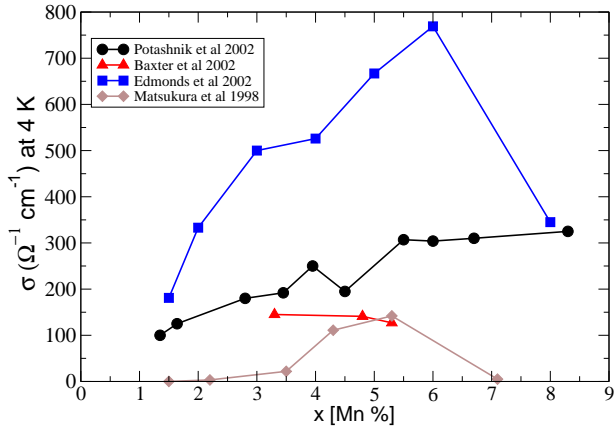}
\caption{Conductivity  $\sigma$ measured  at  $4$ K  or  lower vs.  Mn
percentage  concentration for  (Ga,Mn)As. The  symbols  correspond to:
squares    to    FIX Ref. 24,    circles    to
Ref.   64,         triangles        to
Ref.   65,       and       dimonds       to
Ref. 59.}
\label{rhocomparison}
\end{figure}

The  number  of  research  groups  involved in  the  materials  growth
process,  each  trying a  slightly  different  annealing process,  has
increased over the  past few years and with it  a dramatic increase 
of carrier concentration and conductivity has   taken  place,  as   shown  in
Fig.  \ref{rhocomparison}.   At  the  same  time the  $T_c$  has  also
increased in accordance with the
mean field theory  prediction that $T_c\propto x p^{1/3}$,  where x is
the     Mn     concentration     and     $p$    is     the     carrier
concentration.  The shoulder in $\rho$ observed in 
the most recent optimally annealed samples
near $T_c$ has been partially explained theoretically in terms of 
the variation of the Fermi surface and the transport 
scattering time associated with the ferromagnetic to paramagnetic phase 
transition. \cite{Lopez-Sancho:2003_}
However,   there  is  no  model  at  present  that
reproduces fully the behavior observed in $\rho$ as a function
of temperature near  $T_c$ in the metallic regime for most samples.  
There has also been  theoretical  progress   in  understanding  the  low  temperature
regime\cite{Jungwirth:2002_c}  (Sec.  \ref{boltzman})  and the role
of scattering off magnetic impurities in  the
high             temperature            paramagnetic            regime
(Sec. \ref{jpdfromrho}).\cite{Matsukura:1998_,Omiya:2000_}

\subsection{Scattering off the kinetic-exchange potential in
the paramagnetic regime}
\label{jpdfromrho}

The  contribution from magnetic impurities to the
resistivity  behavior   observed  above  $T_c$  can  be
understood assuming  
scattering from the kinetic-exchange term in the Hamiltonian
\begin{equation}
{\cal H}_{k-e}=J_{pd}\sum_I \delta({\bf r}-{\bf R}_I){\bf S}_I\cdot{\bf s}.
\end{equation}
The corresponding contribution to the resistivity is given 
by\cite{Dietl:1994_,Omiya:2000_}
\begin{equation}
\rho_s=2\pi^2\frac{k_F}{p e^2}\frac{(m^{\ast})^2 J_{pd}^2}{h^3}N_{Mn}
[2\chi_{\perp}(T,B)+\chi_{||}(T,B)],
\end{equation}     
where  $k_F$ is the carrier Fermi wave  vector, $h$  is the  Planck constant,
$N_{Mn}$ the density of substitutional Mn, $m^{\ast}$ 
is the effective mass of the carrier, and $e$
is  the electron  charge.  $\chi_{\perp}=M/B$ and  $\chi_{II}=\partial
M/\partial   B$   are  the   transverse   and  longitudinal   magnetic
susceptibilities. Using transport data, the susceptibility
can  be determined  
from the  Hall resistivity due to the dominant contribution from
the anomalous Hall effect which is proportional to       the
magnetization.\cite{Omiya:2000_} The analysis
of
magnetoresistance  data above $T_c$  gives $J_{pd}=55  \pm 10
{\rm meV nm^{3}}$.  This result is in agreement  with optical MCD data
(see Sec. \ref{optics}). 
We also note that the initial expression used
to   analize   the   magnetoresistance   data   which   replaces   the
susceptibility   factor   by   $[S(S+1)-\langle  {\bf   S}\rangle^2]$,
neglected   the  correlations   between  neighboring   Mn   spins  and
overestimated $J_{pd}$ by a factor  of 3 in disaggrement then with the
MCD measurement of $J_{pd}$. \cite{Matsukura:1998_}

\subsection{Boltzman transport theory of the zero temperature
conductivity in DMS}
\label{boltzman}
The  zero temperature 
conductivity of metallic DMS samples
can be obtained from the effective Hamiltonian (Eq.~\ref{Heff})
and by accounting for disorder scattering
perturbatively.   
The valence band holes
interact  with  randomly  located  spins of  substitutional  Mn
impurities  via  the kinetic-exchange  interaction,  
and  with  randomly  located
ionized  defects and  each other  via Coulomb  interactions.  At zeroth
order, the  interactions are
replaced by  their spatial averages,  so that the  Coulomb interaction
vanishes and  hole quasiparticles  interact with a  spatially constant
kinetic-exchange field. The corresponding mean-field 
Hamiltonian for the  itinerant holes reads
\begin{equation}
{\cal H}_0={\cal H}_{KL}+J_{pd}N_{Mn} S \hat \Omega \cdot\vec s\;,
\label{HMF}
\end{equation}
where  $\hat
\Omega$ is the orientation of fully polarized 
substitutional Mn local moments and   $\vec  s$   is  the
envelope-function hole spin operator. \cite{Abolfath:2001_b} Using the
eigenstates of the Hamiltonian in Eq.~\ref{HMF}, the first order Born approximation of the 
elastic scattering rate, and the 
relaxation-time-approximation solution  to the semiclassical Boltzmann
equation, the diagonal  dc conductivity tensor along one of
the cube edges of the host lattice can be written as:\cite{Jungwirth:2002_c}
\begin{eqnarray}
\sigma_{\alpha\alpha}&=&\frac{e^2}{\hbar V}\sum_{n,k}
\frac{1}{\hbar\Gamma_{n,\vec k}}
\left( \frac{\partial E_{n,\vec k}}{\partial k_{\alpha}}\right )^2
\delta(E_F-E_{n,\vec k}) \; ,
\label{sigma}
\end{eqnarray}
where 
$\Gamma_{n,\vec k}$ 
is the quasiparticle elastic scattering rate, 
$n$ and $\vec k$ are the band and wavevector
indices, 
$E_{n,\vec k}$ are the eigenstates of the Hamiltonian (\ref{HMF}),
and $E_F$ is the Fermi energy.  

 The Born approximation
  estimate of  the transport
weighted  scattering  rate  from  substitutional Mn 
impurities  is given by:
\begin{eqnarray}
\Gamma^{Mn}_{n,\vec k}&=&\frac{2\pi}{\hbar} N_{Mn}\sum_{n^{\prime}}
\int\frac{d\vec k^{\prime}}{(2\pi)^3} 
|M_{n,n^{\prime}}^{\vec k,\vec k ^{\prime}}|^2
\nonumber \\ &\times &\delta(E_{n,\vec k}
-E_{n^{\prime}\vec k ^{\prime}})
(1-\cos \theta_{\vec k, \vec k ^{\prime}})\; ,
\label{gamma}
\end{eqnarray}
with the scattering matrix element,  
\begin{eqnarray}
M_{n,n^{\prime}}^{\vec k,\vec k ^{\prime}}&=&
J_{pd}S
\langle z_{n \vec k}|\hat \Omega\cdot \vec s|
z_{n^{\prime}\vec k ^{\prime}}\rangle\nonumber \\
&-&
\frac{e^2}{\epsilon_{host}\epsilon_0(|\vec k -\vec k ^{\prime}|^2
+q_{TF}^2)}\langle z_{n \vec k}|
z_{n^{\prime}\vec k ^{\prime}}\rangle .
\label{mnelement}
\end{eqnarray}
Here $\epsilon_{host}$ is  the host semiconductor dielectric constant,
$|z_{n   \vec   k}\rangle$   is  the   six-component (multi-component)  
envelope-function
eigenspinor  of  the  unperturbed  Hamiltonian (\ref{HMF}), 
and  the  Thomas-Fermi
screening   wavevector   
$q_{TF}=\sqrt{e^2 DOS(E_F)/(\epsilon_{host}\epsilon_0)}$,
where $DOS(E_F)$ is the density of states at the  Fermi
energy.\cite{Jungwirth:2002_c} This  model incorporates the  fact that
the transport properties of  these materials are not solely determined
by the scattering from  substitutional Mn impurities and allows
explicitly for  scattering from compensating defects,  which have been
seen to  play a  key role in  the resistivity through  the post-growth
annealing                         as                         discussed
earlier. \cite{Ohno:1999_a,Edmonds:2002_b,Ku:2003_,Mathieu:2002_cond-mat/0208411,Yu:2002_a,Edmonds:2004_,Kuryliszyn-Kudelska:2003_cond-mat/0304622}
As-antisite defects are non-magnetic double-donors
and contribute to scattering through a screened charge $Z=2$ Coulomb
potential.  The double-donor  
Mn  interstitials\cite{Yu:2002_a,Edmonds:2002_b}  are
unlikely to be  magnetically ordered and can also  be modeled as a $Z=2$
screened Coulomb potential. \cite{Masek:2002_cond-mat/0201131}

Assuming a  parabolic-band disperssion for  majority heavy-hole states
the  kinetic-exchange scattering contribution  to the  scattering rate
can be estimated  by, $\Gamma_{pd}=(N_{Mn}) J_{pd}^2 S^2 m^{\ast}
\sqrt{2m^{\ast}E_F}/(4   \pi   \hbar^4)$. The   Mn   and
As-antisite  Coulomb scattering  leads to  scattering rate
$\Gamma_{C}$         given        by         the        Brooks-Herring
formula.  \cite{Jungwirth:2002_c} For  (Ga,Mn)As, taking  a heavy-hole
effective  mass $m^{\ast}=0.5m_e$,  $p=0.4$~nm$^{-3}$ and  Mn doping
$x=5$\%,  these  estimates   give  $\hbar\Gamma_{pd}\sim  20$~meV  and
$\hbar\Gamma_{C}\sim 150$~meV.  A full numerical six-band calculations
is consistent  with these estimates,  and   predicts that
the  Coulomb contribution to  the elastic  scattering rate  is several
times  larger  than  the  kinetic-exchange  contribution  for  typical
chemical  compositions.  Note  that hese  estimates give  an immediate
check  on the  assumption  of the  theory  itself, since  even in  the
heavily doped  and compensated (Ga,Mn)As DMS,  the lifetime broadening
of the quasiparticle ($\hbar \Gamma$) is smaller than the valence band
spin-orbit coupling  strength ($\Delta_{so}=341$~meV) and  the typical
Fermi energy.

\begin{figure}
\includegraphics[width=3.3in]{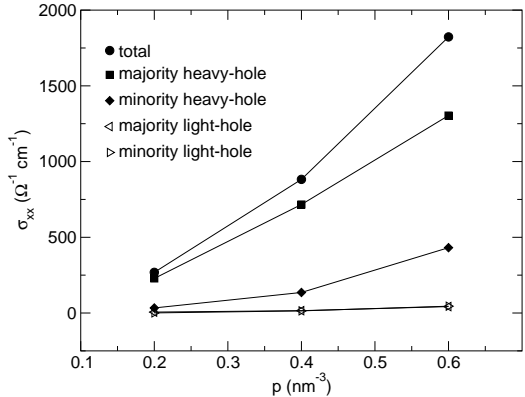}
\caption{Conductivity for measuring current and magnetization directed
along the x-axis  in the plane of the (Ga,Mn)As film  as a function of
the  total  hole density.  These  results  were  obtained for  a  GaAs
semiconductor host doped with 6\%  Mn.  For typical hole densities the
current      is      carried      mostly     by      the      majority
heavy-holes.}
\label{sigmaxx}
\end{figure}

Fig.~\ref{sigmaxx}  shows $\sigma_{xx}$,  calculated  numerically using
the   six-band  Kohn-Luttinger   model   and  Eqs.~(\ref{sigma})   and
(\ref{gamma}), for  a fully strained  Mn$_{0.06}$Ga$_{0.94}$As sample.
The        substrate       --       DMS        lattice       mismatch,
$e_0\equiv(a_{sub}-a_{DMS})/a_{DMS}$, is between  -0.002 and -0.003 in
this case.  \cite{Ohno:1998_a,Edmonds:2003_,Baxter:2002_} The absolute
conductivities predicted by this  model are reasonably consistent with
experiment.~\cite{Potashnik:2002_,Edmonds:2002_b,Baxter:2002_}   The
dissagreement for lower Mn concentrations ($x<4$\%) of the theoretical
conductivities is most likely due to some combination of inaccuracy in
the scattering amplitude estimates, unaccounted sources of scattering,
and,  especially  at  small  $x$,  coherent  scattering  effects  that
eventually lead to localization observed  as an upturn in $\rho$ at the
lowest temperatures.

\begin{figure}
\includegraphics[width=3.3in]{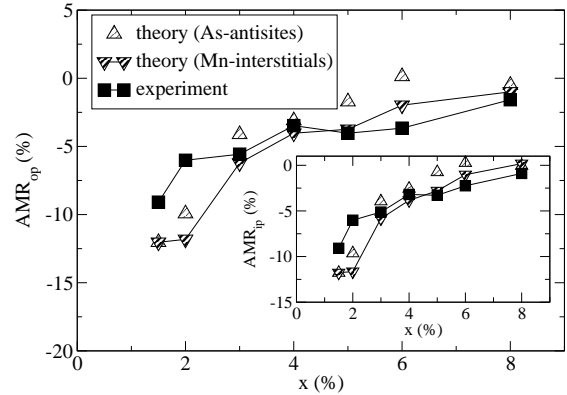}
\caption{Experimental (filled symbols) AMR coefficients and theoretical 
data obtained assuming As-antisite compensation (open symbols) and
Mn-interstitial compensation (semi-filled symbols) 
of $AMR_{op}$ and $AMR_{ip}$. After Ref. 52}
\label{AMR_teor_exp}
\end{figure}

In addition to the AHE (Sec. \ref{AHE}), strong spin-orbit coupling in
the  semiconductor valence  band  leads also  to  anisotropies in  the
longitudinal   transport  coefficients.    In  particular,   when  the
magnetization ${\bf  M}$ is rotated  by applying an  external magnetic
field  stronger  than  the  magneto-crystalline anisotropy  field  the
in-plane                     conductivity                     changes.
\cite{Wang:2002_cond-mat/0211697,Baxter:2002_} Fig.~\ref{AMR_teor_exp}
shows   the    theoretical   and   experimental    AMR   coefficients,
$AMR_{op}=[\sigma_{xx}({\bf   M}\parallel  \hat{z})-  \sigma_{xx}({\bf
M}\parallel  \hat{x})]/\sigma_{xx}({\bf  M}\parallel  \hat{x})  $  and
$AMR_{ip}=[\sigma_{xx}({\bf   M}\parallel  \hat{y})-  \sigma_{xx}({\bf
M}\parallel \hat{x})]/\sigma_{xx}({\bf M}\parallel \hat{x}) $, for the
seven                                                         (Ga,Mn)As
samples.    \cite{Jungwirth:2003_b,Edmonds:2002_b,Baxter:2002_}    Here
$\hat{z}$  is the  growth direction.   Results of  the  two disordered
system   models,  one   assuming   As-antisite  and   the  other   one
Mn-interstitial        compensation,       are        plotted       in
Fig.~\ref{AMR_teor_exp}. As  in the AHE case,  the theoretical results
are able  to account  semi-quantitatively for the  AMR effects  in the
(Ga,Mn)As DMS, with somewhat better agreement obtained for the
model  that  assumes   Mn-interstitial  compensation,  which confirms
indirectly the experimental finding  that the compensating defects are
dominated by Mn-interstitials. \cite{Yu:2002_a} Note that although the
magnitude  of  the  conductivities   tend  to  be  overestimated,  the
magnetotransport  effects  are  relatively insensitive  to  scattering
strength,  reflecting instead  the strong  spin-orbit coupling  in the
valence  band of  the  host  semiconductor as  compared  to the  Fermi
energy.\cite{Jungwirth:2003_b}. 

A large anisotropic magnetoresistance
effect is observed experimentally also in the off-diagonal symmetric
conductivity component ($\sigma_{xy}=+\sigma_{yx}$) when magnetization
direction changes from parallel to $\hat{x}$- or $\hat{y}$-axis, 
where 
$\sigma_{xy}=\sigma_{yx}=0$, 
to a general orientation in the x-y plane tilted from
the two transport measurement axes, 
where $\sigma_{xy}=\sigma_{yx}\neq 0$.\cite{Tang:2003_}
The effect  is called sometimes a 'planar Hall effect' but we emphasize
that the true Hall response requires the symmetry 
$\sigma_{xy}=-\sigma_{yx}$. The anomalous contribution to this asymmetric
off-diagonal transport coefficient is discussed in the following section.  

\subsection{Hall resistivity and the Anomlaous Hall effect}
\label{AHE}

One of the most important characterization tools of magnetic materials
is  the anomalous  Hall effect  \cite{noteonAHE}. The  Hall resistance
$R_{Hall}\equiv \rho_{xy}/d$ 
of a  magnetic thin film is empirically
observed to contain two  distinct contributions. The first arises from
the  normal Hall  effect  contribution which  is  proportional to  the
applied magnetic field, $H$, and the second, called the anomalous Hall
contribution,  is observed  to  be proportional  to the  magnetization
\cite{Chien:1980_}:
\begin{equation}
R_{Hall}=R_0 H+R_S M,
\end{equation}
where $M$ is the magnetization perpendicular to the thin film surface,
and $R_0$ and  $R_S$ are the ordinary and  anomalous Hall coefficients
respectively. The ordinary  Hall effect is due to  the Lorenz force on
the carriers  and is used to  measure the concentration  and nature of
the free  carriers since  $R_0$ is linearly  proportional to  the free
carrier density and its sign  determines whether they are electrons or
holes.  On the  other  hand, the  anomalous  Hall effect  is a  direct
consequence of the  presence of spin-orbit coupling in  the system. In
many  instances,   such  as  the   DMS  for  example,   the  anomalous
contribution is much greater than the ordinary Hall effect, making the
carrier density  hard to  measure since large  magnetic field  must be
used   to  extract   the  linearity   in   the  field   of  the   Hall
resistance. However, the overwhelming of the Hall coefficient by $R_S$
allows one to utilize the anomalous Hall effect as an indirect measure
of  the  magnetization.  Empirically   $R_S$  is  observed  to  depend
quadratically  or  linearly  on  the  diagonal  resistivity,  $\rho$,
depending  on  the  material.  For  example,  for  the  case  of  Fe
\cite{Chien:1980_} and GaMnAs \cite{Edmonds:2003_} the dependance on
the  resistivity is  quadratic.  These two  resistivity dependencies  of
$R_S$ correspond to  different possible 
origins of  the anomalous  Hall effect
which  we discuss  below. However,  in spite  of its  wide use,  it is
surprising how much  confusion has followed over the  past six decades.
Although  much progress has taken place
over the past few years with  the creation and study of new materials,
no  single theory  has been  able to  systematically  and convincingly
explain the constant puzzle of side-jump vs. intrinsic mechanisms (see
below)  although   it  is  at  present  a   feasable  but  challenging
undertaking.

\subsubsection{The embroiled history of the anomalous Hall effect}
Before discussing the theoretical description of
the anomalous  Hall effect  in DMS, it  is appropriate to make an 
excursion to the history of the field
in order to understand clearly the origin
of the controversies and the strengths and weaknesses of the different
arguments put forth by different researchers. Often such summaries
are done in two or three sentences giving an oversimplified view
of the  controversy and the different  points of view,  some of which
have been  challenged  in later literature  but still remain  in the
general folklore.

The  first detailed  theoretical  consideration of  the anomalous  Hall
effect  was  given by  Karplus  and Luttinger  \cite{Karplus:1954_},
where they  considered the problem  from a perturvative point  of view
(with  respect   to  an  applied   electric  field)  and   obtained  a
contribution  to  the Hall  conductivity  in  systems with  spin-orbit
coupled Bloch  states given  by the expression  (Eq. 2.16 and  2.17 in
Ref. ~\onlinecite{Karplus:1954_})
\begin{equation}
\sigma_{xy}=-\frac{2e^2}{\hbar V}\sum_{n,\vec{k}} n_F(E_{\vec{k},n})
{\rm Im}\left\langle \frac{\partial u_{n,\vec{k}}}{\partial k_y}\right.
\left|\frac{\partial u_{n,\vec{k}}}{\partial k_y}\right\rangle,
\label{sigAHEberry}
\end{equation}
where  $n_F$  is  the  Fermi  occupation number  of  the  Bloch  state
$|u_{n,\vec{k}}\rangle$.  This  expression  is  obtained  by  ignoring
impurity scattering (i.e.  clean limit) and can also  be obtained from
the Kubo linear  response theory in this clean  limit (see below). One
of  its   immediate  successes  was  the  prediction   that  $R_S$  is
proportional to  $\rho^2$ in agreement  with many materials,  Fe being
the  primary  example.    Immediately  after Karplus
and Luttinger's pioneer work,  Smit  \cite{Smit:1955_}
considered  the  problem  of  impurity scattering  within  a  Boltzman
formalism using  a model  Hamiltonian of an  electron gas,  i.e. plain
waves,  without   intrinsic  spin-orbit  coupling   and  a  scattering
potential containing both direct and spin-orbit coupled terms:
\begin{equation}
{\cal H}=\frac{\hat{p}^2}{2m}+V(r)+\frac{1}{2m^2 c^2 r}\frac{\partial 
V(r)}{\partial r} \hat{S}_z \hat{L}_z
\label{scatt}
\end{equation}
where the impurity potential is  $V(r)=0$ for $r>R$ and $V(r)=V_0$ for
$r<R$.  Within  the Boltzman formalism,  he obtained a second  type of
contribution to the Hall conductivity due to the asymmetric scattering
from impurities originating from the spin-orbit coupling term. This so
called  skew-scattering   mechanism  predicts,  in   contrast  to  the
intrinsic  effect found  by Karplus  and Luttinger,  that $R_S\propto
\rho$ (i.e. density of scatterers) and dependens on the type and range
of the scattering  potential. It was also noted  that the magnitude of
the effect within  such simplified model was many  orders of magnitude
lower than  the observed  magnitude of  the AHE but,  as it  was shown
later,  the  origin  of  this  deficiency is  the  lack  of  inrtinsic
spin-orbit coupling present in the crystal Bloch states. However, Smit
also critiziced Karplus and Luttinger's result claiming the intrinsic
contribution vanishes. Such claim (see below) is now understood to
be unjustified but has remained  lingering in the AHE literature, generating
much confusion.

Karplus  and Luttinger \cite{Karplus:1954_} 
did not  consider impurity  scattering, hence
omitted  the  skew-scattering contribution  in  their calculation.  To
remedy this, Kohn and  Luttinger \cite{Kohn:1957_} developed a 
consistent (and somewhat
coumbersome)   treatment  to   obtain   the  transport
coefficients  in  the presence  of  spin-orbit  coupling and  disorder
scattering  based on  a density  matrix expansion.  Such  approach is
equivalent to the  Kubo linear response formalism  but technically
more difficult when applied to 
real physical systems. Within this formalism the
expansion is done  to first order in spin-orbit  coupling and impurity
scattering strengths. They obtained,  besides the previous Karplus and
Luttinger  result (Eq.  135 in  Ref.  ~\onlinecite{Kohn:1957_}, however
note   that    it   has   the   wrong   sign    later   corrected   in
Ref. ~\onlinecite{Luttinger:1958_}),  formal expressions for additional
contributions due to impurity scattering. Applying this formalism to a
model  with   uncorrelated  impurity  potential   and  small  scattering
strength,  Luttinger \cite{Luttinger:1958_}  obtained,
besides the Smit's skew-scattering term, the same  final
expression as Karplus and Luttinger  {\em but} with the opposite sign
(Eq. 3.26 in  Ref. ~\onlinecite{Luttinger:1958_}). Somehow, in addition
to  the  intrinsic Karplus and Luttinger term,  
there  seemed  to  be an  additional
contribution from scattering {\em in the high mobility limit} giving a
contribution identical to the intrinsic  one but with a factor of $-2$
in front  of it, hence  the flip of  sign of the final  expression. No
explanation was given of why  such a scattering contribution {\em does
not} depend in  any way on the scattering potential  but simply on the
electronic    structure   of    the   system.     However,   Luttinger
\cite{Luttinger:1958_}  showed   inequivaquely  that  the  cancelation
argued by Smit does not  take place. In a slightly different formalism
Adams  and Blouht \cite{Adams:1959_}  agreed with  Luttinger's results
and  pointed  out  that  Smit's  error occurs  from  an  inconsistency
involving  a  change of  representation  of  his  density matrix  when
calculating the current.

A  few   years  later   Berger  \cite{Berger:1970_}  made   his  first
contribution to  the problem  introducing the idea  (contained withing
Luttinger's  formalism)  of  the  side-jump scattering  mechanism.  He
considered the problem of  wave-paket scattering off an impurity potential
as in  Eq. \ref{scatt},  showing that in  addition to  skew scattering
there exists a side-step type of scattering. This so called side-jump,
as  in the  Luttinger  formalism, was  independent  of the  scattering
potential,  leaving one  with the  strange notion  that  an exterinsic
scattering mechanism resulted in  an intrinsic property. After putting
this contribution in a Boltzman type of formalism he obtained
a result  which was four orders  of magnitde lower  than the Karplus
and Luttinger intrinsic
result. He then  argued that the intrinsic spin-orbit  coulping of the
system increases this side-jump scattering by such factor. In his case
the sign of  the effect is not directly considered  and his result, as
already  mentioned,   is  independent  of   the  scattering  potential
itself. A similar type of formalism  was put forth by Lyo and Holstein
\cite{Lyo:1972_}  claiming  similar  results  but ignoring,  for  some
unknown reason, intrinsic  non-vanishing contributions of Karplus
and Luttinger.

It  is then  that Leroux-Hugon  and  Ghazali \cite{Leroux-Hugon:1972_}
pointed out  the resolution of the  missing order of  magnitude of the
skew-scattering. By  looking simply  at scattering of  Bloch electrons
from ionized  impurities $V_{ion}(r)$ alone and going  to second order
in the  Born approximation within  the collision term of  the Boltzman
transport theory, they obtained  a skew-scattering contribution with a
much higher magnitude than the one coming from the spin-orbit coupling
of the impurities themselves. Hence,  it became evident that it is the
presence of  {\em intrinsic}  spin-orbit coupling through  the crystal
potential which is ultimately responsible for the magnitude of all the
anomalous Hall effect contributions observed and a system with weak or
no  intrinsic  spin-orbit  coupling  will  not  exhibit  a  detectable
contribution to the $R_{Hall}$.

To elucidate  this confusing and farraginous  atmosphere, Nozieres and
Lewiner \cite{Nozieres:1973_},  following the formalism  introduced by
Fivaz  \cite{Fivaz:1969_}  in   terms  of  effective  Hamiltonians  of
semiconductors,   developed  a   theory  applicable   to   narrow  gap
semiconductors  equivalent to  Luttinger's  \cite{Luttinger:1958_} but
with the final  result in being more physically  transparent. Going to
linear  order  in  impurity  scattering and  spin-orbit  coupling  and
combining  them  with  a  Boltzamn  type of  approach,  they  obtained
different  contributions to  the Hall current  and different  
behavior when
considering opposite orders of limits of $\omega\rightarrow 0$
and $\tau\rightarrow\infty$, where  $\omega$ is
the  frequency  of  the  applied  electric field  and  $\tau$  is  the
quasiparticle lifetime
(see  Table  I   and  Eqs.  59,  60a,   and  60b  in
Ref.   ~\onlinecite{Nozieres:1973_}).  Besides   the   skew  scattering
contribution which depends on the type of scattering potential, in the
limit  $\omega\tau>>1$ ($\tau\rightarrow\infty$ first),  
considered  by  some  authors to  be  the  weak
scattering   limit   (see   Refs.  ~\onlinecite{Chazalviel:1975_}   and
~\onlinecite{Dietl:2003_cond-mat/0306484}),  the  result  is  that  of
Karplus and Luttinger.\cite{Karplus:1954_} On the other hand, for 
$\omega\tau<<1$ ($\omega\rightarrow 0$ first),
the opposite sign expression of Luttinger \cite{Luttinger:1958_}
is recovered. 
One must note however that
this result seems  to be very specific to the  model (single band) and
to the meaning of the
limit $\tau\rightarrow \infty$ as admitted by the authors themselves.~\cite{Nozieres:1973_,Luttinger:1958_}

After this Berger  and Smit exchanged salvos in  a series of confusing
comments\cite{Smit:1973_,Berger:1973_,Smit:1978_,Berger:1978_}       in
which Smit's old argument, that there  is no such a thing as side-jump
or intrinsic contribution, seems to  have been finally put to rest. It
was after    this   that    Chazalviel    \cite{Chazalviel:1975_},   using
Nozieres-Lewier-Luttinger\cite{Luttinger:1958_,Nozieres:1973_}      and
Leroux-Hugon  and  Ghazalis  \cite{Leroux-Hugon:1972_}  results,
attempted to compare phenomenologically the AHE theory
to n-InSb and n-Ge experimental
data with some success. Within such models, the relative importance of
the  side-jump/intrinsic  contribution $\sigma_{aH}^{sj/int}$  (which
are the same  magnitude but of opposite sign  depending on the limits
\cite{Nozieres:1973_,Chazalviel:1975_,Dietl:2003_cond-mat/0306484}) to
the  skew-scattring contribution  $\sigma_{aH}^{sk}$  depends on  the
nature   of   the  scatterers   (only   through  the   skew-scattering
dependence).   For  scattering   from  ionized   impurities   one  has
\cite{Leroux-Hugon:1972_,Chazalviel:1975_}
\begin{equation}
\left|\frac{\sigma_{aH}^{sj/int}}{\sigma_{aH}^{sk}}\right|= 
\frac{c N}{p r_s k_F l},
\label{sjtoss_ion}
\end{equation} 
with $N/p$ being  the ratio of the total  number of ionized impurities
and carrier density, $r_s$ is  the average distance between carriers in
units of Bohr radious, $l$ is  the mean free path, and $c\sim 10$, varying
slightly with scattering length.  For short range scattering potential
considered by Luttinger  \cite{Luttinger:1958_} and Nozieres and Lewiner
\cite{Nozieres:1973_}, $V(\vec{ r})=V_0\delta(\vec{r}-\vec{r_i})$:
\begin{equation}
\left|\frac{\sigma_{aH}^{sj/int}}{\sigma_{aH}^{sk}}\right|=\frac{3}
{\pi |V_0| D(E_F) k_F l},
\label{sjtoss_short}
\end{equation}
where $D(E_F)$ is the density of  states at the Fermi energy and $k_F$
is the Fermi wave-vector. These  estimates can be a useful first guess
at which mechanism dominates in  different materials but one must keep
in mind the simplicity of the models used to estimate such ratio.

In  spite  of   the  theory  being  incomplete,  and   in  some  cases
self-contradictory and puzzling, the theoretical developments remained
somewhat dormit until recent  interest in new ferromagnetic and spintronic
materials. These emerging field have demonstrated the
need for a better  theoretical understanding of the AHE
which has constantly  been used to characteize ferromagnetic
materials.

We emphasize that there is,  at this stage,  no controversy regarding
the skew-scattering contribution. In most materials such contribution,
although  present, is minor  and in  most ferromagnetic  materials the
quadractic  dependence  of  $R_S$  is  observed  even  at  the  lowest
temperatures. However,  if somehow,  one could artificially  turn down
$\rho$ one would eventually reach  a regime where such contribution is
dominant, a  fact which  very few people  dispute since the  origin of
such contribution  is rather transparent and  reasuringly extrinsic in
nature.  However, the  flip  of  sign of  the  expressions derived  by
Luttinger\cite{Luttinger:1958_}         and        Nozieres        and
Leweir\cite{Nozieres:1973_} within  the simple  one band model  in the
different order  of limits of $\omega\tau $,  remains quite disturbing
and a source  of continuous debates. After all,  these results are only
obtained  in a  limit where,  supposedly, the  skew  scattering would
always dominate. As we will see below, it seems that in many materials
experimental comparisons, the ultimate test of validity of any theory,
are    more    consistent     with    the    intrinsic Karplus
and Luttinger \cite{Karplus:1954_}   contribution
than the  reversed  sign one of Luttinger \cite{Luttinger:1958_};  
the  sign of  the
effect being a simple thing to check.

A  possible resolution to  the problem  would be  to perform  the Hall
conductivity  calculation within the  Kubo linear  response formalism,
treating disorder  and spin-orbit coupling  on an equal  footing. This
can be achieved  by starting from the Dirac  representation and taking
the  weak relativistic  limit  after treating  disorder  in the  usual
fashion.  In  this way,  any  spin-orbit  coupling contribution  comes
naturally from the Dirac representation and therefore is automatically
taken into  account. On the other  hand, if one starts  from the Pauli
Hamiltonian,  an inmediate question  arises: which  vertex corrections
capture the different contributions to the anomalous Hall effect? Such
an   approach   was   taken   recently   by Crepieux, Dugaev, and  Bruno.~\cite{Crepieux:2001_b,Crepieux:2001_a,Dugaev:2001_} 
The work considers, however,  a free  electron gas model rather  than a
crystaline  enviroment,   hence  ignoring  the   intrinsic  spin-orbit
coupling effects. In spite  of this simplification, the authors 
obtained many
useful  findings.  They where  able  to  pinpoint which  diagrams
correspond  to the  skew and  the  side-jump scattering  in the  Pauli
Hamiltonian by  identifying them  from the corresponding  ones arising
naturally from  the Dirac formalism. Notably,  these diagrams within
the Pauli Hamiltonian,  are not the standard ones  that one would take
into  account  naturally.  Their  main   result,  Eq.  50  and  51  in
Ref.   ~\onlinecite{Crepieux:2001_b,Dugaev:2001_},  is   in  reasonable
agreement with Berger's simpler treatment of this free electron
gas problem.

Taking  into consideration  the  confusion generated  by the  possible
relevance of  the side-jump scattering  mechanism, several researchers
have chosen to focus instead on the original intrinsic contribution to
the  AHE  proposed by  Karplus  and Luttinger 
\cite{Karplus:1954_} and
ignore impurity scattering all  together or simply include its effects
through  the  Born  approximation  which introduces a  finite
quasiparticle lifetime. In DMS, e.g., this approach is partly motivated
by the strong  intrinsic spin-orbit coupling
in the host valence bands that makes much of the above theoretical
discussions inapplicable 
since they relied on the perturbation treatement of the spin-orbit
coupling. One of the main attraction of the intrinsic AHE theory is
the  ability  to  do   calculations  in  models  with   realistic
electronic band-structures.  This approach  has been used, e.g., to analyze
the AHE in layered 2D  ferromagnets such as SrRuO$_3$, in pyrochlore
ferromagnets
\cite{Onoda:2002_,Fang:2003_,Taguchi:2001_,Shindou:2001_}, in         the
collosal  magnetoresistance manganites, \cite{Ye:1999_} in Fe,
\cite{Yao:2004_} in DMS,\cite{Jungwirth:2002_a}  
and  as a natural starting point to  address infrared magneto-optical effects such
as the Kerr and Faraday effects \cite{Sinova:2003_} (see Sec. V).  The
application   of   this   approach   to  these   different   materials
and their  succesful comparison to  experiments 
\cite{Taguchi:2001_,Onoda:2002_,Jungwirth:2002_a,Jungwirth:2003_b,Edmonds:2003_,Edmonds:2002_a,Fang:2003_,Yao:2004_}
is perhaps  the most
pogniant  criticism  to  the  old  theories  regarding  the  side-jump
scattering as fundamental.

\subsubsection{Anomalous Hall effect in DMS}

The anomalous Hall effect in DMS  has been one of the most fundamental
characterization tools  since it is  the simples way of  detecting the
ferromagnetic state of the system at a given temperature. The original
discovery   of  ferromagnetism   in   (III,Mn)V  seminconductors   was
established by measuring the AHE in both the high and
low temperature regimes.~\cite{Ohno:1992_} The comparison
with  remenant magnetization measurements  using a  SQUID magnetometer
confirmed, e.g., that AHE measurement can be used to determine
ferromagnetic  
critical temperature.~\cite{Ohno:1992_}

\begin{figure}
\includegraphics[width=3.3in]{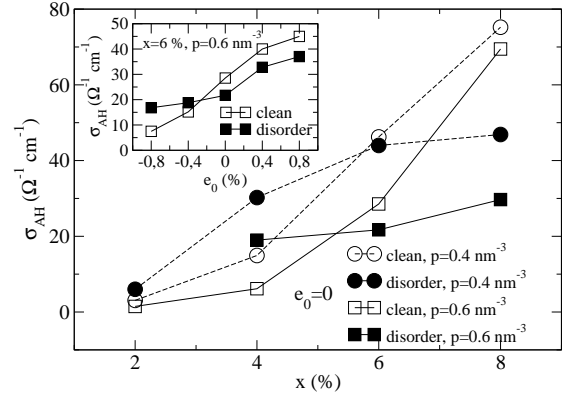}
\caption{Theoretical anomalous Hall conductivity of Mn$_x$Ga$_{1-x}$As
DMS calculated  in the clean  limit (open symbols) and  accounting for
the  random  distribution of  Mn  and  As-antisite impuriries  (filled
symbols), after Ref.  52.}
\label{AHE_teor}
\end{figure}

Recently  Jungwirth, Niu, and MacDonald \cite{Jungwirth:2002_a} reintroduced
the  original Karplus  and Luttinger  Eq. \ref{sigAHEberry}, pointing
out that the intrinsic contribution to the AHE is proportional to
the Berry phase acquired by a quasiparticle wave function upon traversing
closed path on the spin-split Fermi surface. They
applied the theory to metallic  
(III,Mn)V materials using  both the 4-band  and 6-band 
$\vec{k}\cdot\vec{p}$ description of
the  valence band electronic 
structure\cite{Abolfath:2001_b} and obtained results
in a quantitative agreement 
with  the  experimental data in (Ga,Mn)As and (In,Mn)As DMS.
In  DMS systems,  the
estimate given in  Eq. \ref{sjtoss_ion} gives a ratio  of intrinsic to
skew scattering contribution  of the order of 50,  hence the intrinsic
contribution      in     these      systems     is     likely     to
dominate.\cite{Dietl:2003_cond-mat/0306484}  Consistently,
Edmonds
{\it    et   al.}\cite{Edmonds:2002_a}  found that 
in
metallic DMS systems  $R_s\propto \rho^2$. In    a   follow    up   work
\cite{Jungwirth:2003_b}  a  more  careful  comparison  of  the  theory,
reformulated within  the Kubo formalism,  was done in order  to account
for   finite  quasiparticle  lifetime   effects  important   only  for
quantitative  but  not qualitative  comparison  with the  experimental
data.  Within  the  Kubo   formalism  the  dc  Hall  conductivity  for
non-interacting quasiparticles is given by
\begin{eqnarray}
\sigma_{\rm xy}&=&\frac{i e^2\hbar}{m^2 }\int \frac{d\vec{k}}{(2\pi)^3}
\sum_{n\ne n'}
\frac{f_{n',\vec{k}}-f_{n,\vec{k}}}{E_{n\vec{k}}-E_{n'\vec{k}}} 
\nonumber\\&&\times
\frac{\langle n \vec{k}|\hat{p}_x|n'\vec{k}\rangle\langle 
n'\vec{k}| \hat{p}_y|n \vec{k}\rangle}
{E_{n'\vec{k}}-E_{n\vec{k}}+i\hbar\Gamma}.
\label{sigAHEKubo}
\end{eqnarray}
Looking at the real part of the dc Hall conductivity in the clean limit 
($\hbar\Gamma\rightarrow 0$), 
the delta-function contribution from the denominator 
vanishes due to the Fermi factor differences and we obtain
\begin{eqnarray}
\sigma_{\rm xy}&=&\frac{e^2\hbar}{m^2 }\int \frac{d\vec{k}}{(2\pi)^3}
\sum_{n\ne n'}
(f_{n',\vec{k}}-f_{n,\vec{k}}) \nonumber\\&&\times
\frac{{\rm Im}[\langle n' \vec{k}|\hat{p}_x|n\vec{k}\rangle\langle 
n\vec{k}| \hat{p}_y|n' \vec{k}\rangle]}
{(E_{n\vec{k}}-E_{n'\vec{k}})^2}.
\label{sigAHEKubo2}
\end{eqnarray}
Realizing  that the  dipole matrix  elements considered  above  can be
written   as    $\langle   n'   \vec{k}|\hat{p}_{\alpha}|n\vec{k}\rangle=
(m/\hbar)\langle                  n'                  \vec{k}|\partial
H(\vec{k})/\partial_\alpha|n\vec{k}\rangle$, Eq. \ref{sigAHEKubo2} can
be shown  to be equivalent  to Eq. \ref{sigAHEberry}.  However, within
the Kubo formalism,  it is straightforward to  account for the finite
lifetime  broadening  of the  quasiparticles  within  the simple  Born
approximation by  allowing $\Gamma$ above  to be finite.   The effective
lifetime   for  transitions  between   bands  $n$   and  $n^{\prime}$,
$\tau_{n,n^{\prime}}\equiv 1/\Gamma_{n,n'}$,  can be  calculated by
averaging  quasiparticle  scattering  rates  calculated  from  Fermi's
golden rule including both screened Coulomb and exchange potentials of
randomly  distributed substitutional  Mn and  compensating  defects as
done      in      the      dc     Boltzman      transport      studies
\cite{Jungwirth:2002_c,Sinova:2002_}.   In   Fig.~\ref{AHE_teor}   the
compensation is assumed to be due entirely to As-antisite defects. The
valence  band  hole   eigenenergies  $E_{n\vec{k}}$  and  eigenvectors
$|n\vec{k}\rangle$  in Eqs.~(\ref{sigAHEKubo})-(\ref{sigAHEKubo2}) are
obtained  by solving  the six-band  Kohn-Luttinger Hamiltonian  in the
presence  of  the   exchange  field,  $\vec{h}=N_{Mn}S  J_{pd}\hat{z}$
\cite{Abolfath:2001_b}.  Here $N_{Mn}=4x/a_{DMS}^3$  is the Mn density
in the Mn$_x$Ga$_{1-x}$As epilayer  with a lattice constant $a_{DMS}$,
the  local  Mn  spin  $S=5/2$,  and  the  exchange  coupling  constant
$J_{pd}=55$ meV nm$^{-3}$.

Fig.~\ref{AHE_teor}  demonstrates  that  whether  or not  disorder  is
included, the  theoretical anomalous Hall conductivities  are of order
10~$\Omega^{-1}$~cm$^{-1}$  in  the (Ga,Mn)As  DMS  with typical  hole
densities, $p\sim 0.5$~nm$^{-1}$, and Mn concentrations of several per
cent. On a quantitative level, disorder tends to enhance $\sigma_{AH}$
at low  Mn doping and suppresses  AHE at high  Mn concentrations where
the quasiparticle broadening due to disorder becomes comparable to the
strength of the exchange  field. The inset in Fig.~\ref{AHE_teor} also
indicates that  the magnitude of the  AHE in both  models is sensitive
not only  to hole  and Mn densities  but also to  the lattice-matching
strains     between    substrate     and    the     magnetic    layer,
$e_0=(a_{substrate}-a_{DMS})/a_{DMS}$.

\begin{figure}
\includegraphics[width=3.3in]{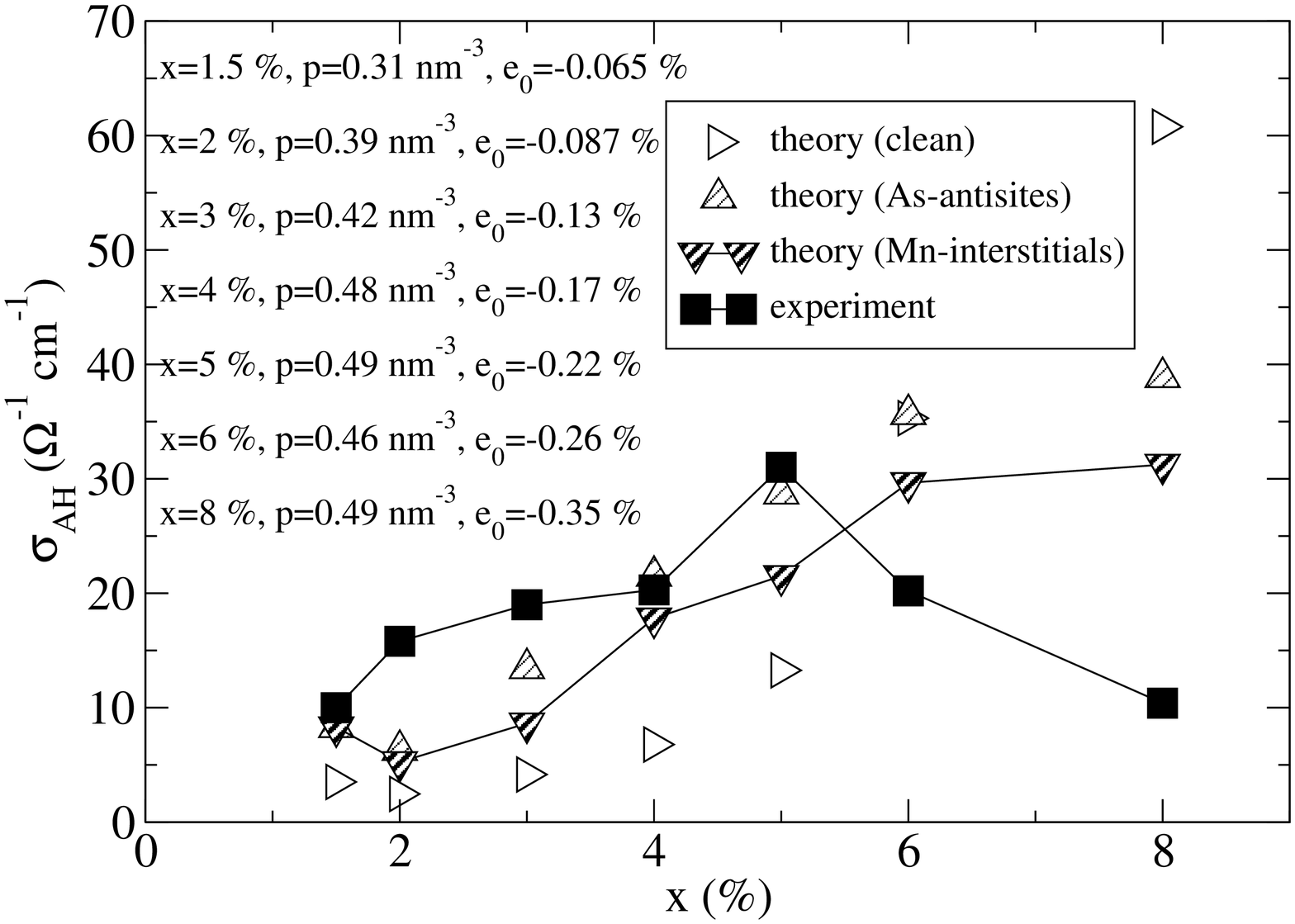}
\caption{Comparison between experimental and theoretical anomalous Hall
conductivities, after Ref. 52.}
\label{AHE_teor_exp}
\end{figure}

A systematic comparison between  theoretical and experimental AHE data
is   shown  in  Fig.~\ref{AHE_teor_exp}.   \cite{Jungwirth:2003_b}  The
results  are plotted  vs.  nominal Mn  concentration  $x$ while  other
parameters  of the  seven samples  studied  are listed  in the  figure
legend.   The measured  $\sigma_{AH}$ values  are indicated  by filled
squares; triangles are theoretical results obtained in the clean limit
or  for  a  disordered  system  assuming  either  the  As-antisite  or
Mn-interstitial compensation  scenario.  In general,  when disorder is
accounted for,  the theory  is in a  good agreement  with experimental
data over  the full  range of studied  Mn densities from  $x=1.5\%$ to
$x=8\%$.   The   effect   of   disorder,  especially   when   assuming
Mn-interstitial  compensation, is particularly  strong in  the $x=8\%$
sample   shifting  the  theoretical   $\sigma_{AH}$  much   closer  to
experiment  compared   to  the  clean  limit   theory.  The  remaining
quantitative  discrepancies between  theory and  experiment  have been
attributed  to the resolution  in measuring  experimental hole  and Mn
densities.\cite{Jungwirth:2003_b}

We conclude this section by  mentioning other studies of the anomalous
Hall  effect in  regimes  and  geometries which  we  have not  focused
on here.  A study  within the  non-metallic regime  or  hopping transport
regime, not covered  in any detailed within this  review, has been done
by Burkov {\it et al.}\cite{Burkov:2003_} following the ideas utilized
in  the mettalic  Berry's phase  approach but  generalized  to hopping
transport regime. Their results are in good agreement with experimental
data in these  regime which encompass samples with less than
1\% doping.   At temperatures above $T_c$, a theory based on semiclassical
Boltzman  transport  has  suggested  the presence  of  anomalous  Hall
voltage noise within this regime which could be a further test of the
intrinsic
AHE model.\cite{Timm:2003_cond-mat/0309547} 

A
recent theoretical
study by  Bruno {\it et al.} \cite{Bruno:2003_cond-mat/0310522}
has  suggested  an alternative experimental system 
to  test the  topological Hall
effect   arising  from   the   Berry  phase   similar  to   Eq.
\ref{sigAHEberry}.  Here  the difference  is  that  the  system has  a
spatially  varying magnetization.  Upon a  local  gauge transformation
which   makes  the   quantization  axis   oriented  along   the  local
magnetization  direction, the  problem  becomes that  of a gauge  vector
potential corresponding to monopole  fields. Such a system
can be created by sandwiching  a thin film (II,Mn)VI semiconductor and
a  lattice  of iron  nano-rods  oriented  perpendicular  to the  doped
semiconductor epilayer.  Future experiments might show whether such a
device can be grown efficiently and the theory tested.

\section{Optical properties of DMS}
\label{optics}

Optical properties  are among the key  probes 
into the electronic  structure of
materials and, in the case  of semiconductors, can also be utilized to
change   its  properties   by  photo-carrier   generation   and  other
means. Given the wide energy range that optical probes can
attain, many different physics can  be addressed by exploring the full
spectrum   of   phenomena   including  the visible  and   infrared
absorption,\cite{Singley:2002_,Singley:2003_,Chapman:1967_,Hirakawa:2002_,Hirakawa:2001_,Nagai:2001_,Szczytko:1999_a}
magneto-optical     effects      such     as  the   magnetic     circular
dichroism,\cite{Szczytko:1996_,Beschoten:1999_,Szczytko:1999_a,Ando:1998_}
Ramman          scattering,\cite{Seong:2002_,Limmer:2002_,Sapega:2001_}
photoemission,\cite{Okabayashi:1999_,Okabayashi:2002_,Okabayashi:2001_,Asklund:2002_}
and                                                           cyclotron
resonance.\cite{Khodaparast:2003_cond-mat/0307087,Sanders:2003_cond-mat/0304434,Mitsumori:2003_cond-mat/307268}
We focus in this review on the first two items of this long list.  The
motivation for  the study  of optical properties  of materials  is two
fold: first, it gives a consistancy check on the different theoretical
models and can be used to measure directly phenomenological parameters
such           as            the           exchange           coupling
$J_{pd}$,\cite{Ando:1998_,Beschoten:1999_} and second, 
it  can  be utilized  in applications such as 
magneto-optical memories, optical
isolators,         and         circulators         for         optical
communications.~\cite{Sugano:2000_}

\subsection{visible and infrared absorption}

The  simplest  and  most  direct  of  the  optical  effects  is  light
absorption,  which probes   the electronic  structure through
electron excitations  between different bands  or impurity
states. An  important measurement in semiconductors is  the band edge
absorption,
usually obtained  in the visible frequency range.  This visible regime
has been exploited more often using absorption from polarized light in
order  to  measure magneto-optical  effects  such  as the MCD and  we  will
postpone its  disucssion to Sec.  \ref{MCD}. However, in  the infrared
regime, many  of the meV  physics is revieled and  several remarkable
experimental    features     have    been    observed     in    recent
experments. \cite{Singley:2002_,Singley:2003_,Hirakawa:2002_,Hirakawa:2001_,Nagai:2001_,Szczytko:1999_a}
These experiments, performed in  thin film geometries, exhibit several
common  phenomena: (i)  a non-Drude  behavior in  which the  conductivity
increases with  increasing frequency in  the interval between  $0$ meV
and $220$  meV, (ii)  a broad absorption  peak near $220-260$  meV that
becomes  stronger  as  the  samples   are  cooled,  and  (iii)  a  broad
featureless absorption between the  peak energy and the effective band
gap energy which tends to increase at the higher frequencies.

In order to understand these features, two theoretical approaches have
been employed mostly. The first
is  the semi-phenomenological effective  Hamiltonian model that focuses on
the multi-band  nature and on the
intrinsic spin-orbit coupling present  in the host
semiconcutor  in  order  to quantitatively
understand these materials and is able to capture  many
of the  observed infrared properties. 
\cite{Sinova:2002_,Yang:2003_,Sinova:2003_}
The  second approach, used  in the  context of  lattice models,  emphasizes the
connection  with the localization  physics present  in the  low doping
regime and simplifies the electronic structure by assuming a single-band,   
seeking  a  more   qualitative  rather   than  quantitative
understanding  of these  materials  in the  regime  of interest.  Both
models are important to attain a full understanding of DMS optical
properties and we review the results of each in turn.

\subsubsection{Multiband effective Hamiltonian approach}
\label{absorb}
\begin{figure}
\includegraphics[width=3.2in]{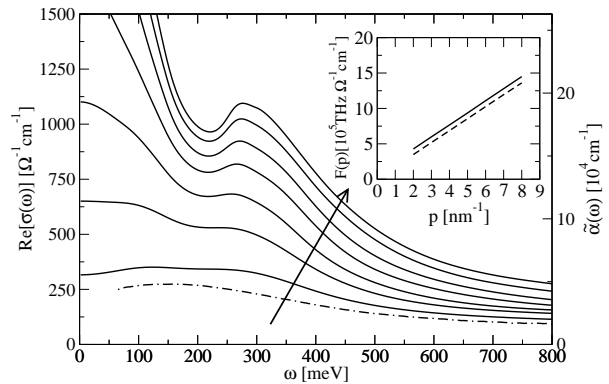}
\caption{Optical    conductivity   ${\rm    Re}[\sigma(\omega)]$   and
absorption coefficient  $\tilde{\alpha}(\omega)$ for carrier densities
from $p=0.2$ to $0.8 {\rm  nm}^{-3}$ in the direction indicated by the
arrow,  for  Ga$_{0.95}$Mn$_{0.05}$As.   The  dot-dashed  line  is  an
experimental absorption curve for a  sample with a Mn concentration of
4\%  obtained from Ref.  {Hirakawa:2002}. The  inset shows
the spectral weight  $F(p)$ evaluated with a frequency  cut-off of 800
meV including disorder (dashed) and  in the clean limit (solid). After
Ref. 45}
\label{fig3}
\end{figure}

Within the effective Hamiltonian  model (Eq. \ref{Heff}), the $220$ meV
peak   can   be    attributed   to   inter-valence-band   transitions,
\cite{Sinova:2002_}   rather   than   to   transitions   between   the
semiconductor  valence band  states  to a  Mn  induced impurity  band,
although a combination of  these contributions will always be present.
In thin  film absorption  measurements, for infrared  wavelengths much
larger than the  width of the film, the real  part of the conductivity
is related to the absorption coefficient by
\begin{equation}
\tilde{\alpha}(\omega)=2\frac{{\rm Re[}\sigma(\omega)]}{Y+Y_0}\,\,,
\end{equation}
where  $Y$ and $Y_0$  are the  admittances of  the substrate  and free
space,  respectively.  For  shorter wavelengths  multiple  reflections
within the  film must be taken  into account to obtain  an estiamte of
the     absorption    coefficient.      The     conductivity    tensor
$\sigma_{\alpha,\beta}(\omega)$  at $T=0$ can  be evaluated  using the
standard Kubo  formula for non-interacting  quasi-particles.  Disorder
is taken into  account within the Born approximation  by including the
lifetime broadening of  quasiparticle spectral functions in evaluating
the Kubo formula. The effective lifetime for transitions between bands
$n$   and  $n^{\prime}$,   $\tau_{n,n^{\prime}}$,  is   calculated  by
averaging  quasiparticle  scattering  rates  calculated  from  Fermi's
golden rule including both  screened Coulomb and exchange interactions
as in Sec.  \ref{boltzman}. Fig.~\ref{fig3} shows the ac conductivities
calculated for ${\rm Ga}_{.95}{\rm Mn}_{.05}$As at a series of carrier
densities.\cite{Sinova:2002_}

An imporant concept in optical absorption spectra is the f-sum rule:
\begin{equation}
F \equiv \int_0^{\infty} d\omega {\rm Re}[\sigma_{xx}(\omega)]=
\frac{\pi e^2}{ 2V} \sum_{\alpha} f_{\alpha} \;\;
\langle \alpha | \frac{\partial^2 H_{KL}} {\hbar^2 \partial k_x^2} | 
\alpha\rangle.
\label{eq:realfsumrule} 
\end{equation}
In  this equation  $f_{\alpha}$ is  a quasiparticle  Fermi  factor and
$\partial^2 H_{KL}/  \hbar^2 \partial k_x^2$ is the  $xx$ component of
the ${\bf k.p}$ model inverse  effective mass operator.  This sum rule
is completely independent of the weak-scattering approximations but it
is necessary  to choose  an upper cut-off  for the  frequency integral
which creates a small uncertainity in the optical effective mass, 
defined
by $F=\pi  e^2 p  / 2 m_{opt}$.   As illustrated  in the inset  of Fig
\ref{fig3}, for  the case of  cut-off frequency $\hbar\omega_{max}=800
{\rm  meV}$, the  f-sum rule  values of  $F$ evaluated  from  our weak
scattering theory are  accurately linear in $p$ over  the entire range
of  relevant  carrier  densities.   Disorder  does have  a  small  but
measurable effect on $m_{opt}$  as illustrated in Fig.~\ref{fig3}. The
optical masses for  GaAs, InAs, and GaSb DMS  ferromagnets with a 800,
400, and 700  {\rm meV} cutoffs are 0.25-0.29  m$_e$, 0.40-0.43 m$_e$,
and  0.21-0.23  m$_e$  respectively,  the  extremes  of  these  ranges
corresponding to  the clean (lower)  and disordered (upper)  limits of
our model.\cite{Sinova:2002_}

Free carrier concentration is difficult to determine accurately in DMS
systems  because  the   anomalous  Hall  contribution  overwhelms  the
oridnary  Hall coefficient  and, in  many instances,  a  high magnetic
field is required.  Hence, the f-sum rule can be  used as an alternate
tool  to measure the  free carrier  concentration in  metallic systems
where the assumed electronic structure  may be a good approximation so
$m_{opt}$  is given  correctly by  the estimates  above. On  the other
hand, accurate estimates of the  carrier concentration $p$ can be used
to       test       the       theory       as       well.       Recent
experiments\cite{Singley:2002_,Singley:2003_} in as-grwon samples have
been interpreted  in both ways  without clear distinction  between the
effective Hamiltonian  picture or the  impurity band picture  since no
measurements  of the  carrier concentration  where  available. Further
measurements in  the most metallic samples  will serve as  test of the
simplifying assumptions within this model.

\begin{figure}
\includegraphics[width=3.4in]{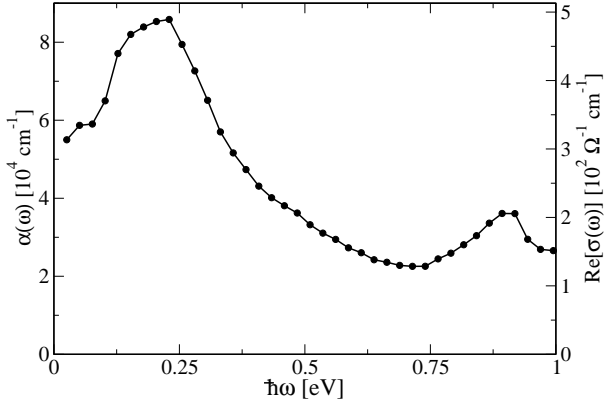}
\caption{Absorption  and conductivity  of a  metallic  sample computed
using a Luttinger Kohn  Hamiltonian with disorder. Here  
$p=0.33$~nm$^{-3}$, $n_{Mn}=1$~nm$^{-3}$ ($x\approx 4.5$\%). After 
Ref. 46}
\label{yang2}
\end{figure}

One prominent feature that is  at odds with experimental data from the
above model calculations is the  relative magnitude of the $\omega \to
0$ conductivity and the $220$ meV conductivity peak. The reason is the
impossibility of the model to account for weak and strong localization
effects from multiple-scattering effects invariably present in the DMS
materials  and  which suppress  te  low-frequency conductivity.  Exact
diagonalization  calculations still  within the  effective Hamiltonian
model  in a  finite  system  size but  treating  the disorder  effects
exactly  can  fix  this  shortcoming  as  shown  in  Fig.  \ref{yang2}
.\cite{Yang:2003_} In  this calculation  the f-sum rule  value differs
from  its Born  approximation  value  by less  than  10\% for  typical
metallic carrier densities, hence justifying the use of $m_{opt}$ as a
measuring tool of the carrier concentration $p$.

\subsubsection{Lattice models and dynamical mean field theory approach}

Within  the  lattice   models,  dynamical  mean-field-theory  studies,
\cite{Hwang:2002_,Alvarez:2003_} for a  single-band model that neglect
the  spin-orbit coupling  and the  heavy-light degeneracy  of  a III-V
semiconductor  valence band,  have sown  that  non-Drude impurity-band
related  peaks in  the frequency-dependent  conductivity occur  in DMS
ferromagnet models when the  strength of exchange interaction coupling
is comparable to the   band width.  The starting Hamiltonian of
these theories is the generalized Kondo lattice model
\begin{eqnarray}
H=H_{host}   &-&   \sum_{i,\alpha,\beta}J\hat{\bf   S}_{i}\cdot   \psi
_{\alpha }^{ \dag}(R_i)\vec{\sigma}_{\alpha \beta } \psi_{\beta }(R_i)
\nonumber \\ &+& W \psi^{\dag}_{\alpha }(R_i) \psi_{\alpha }(R_i),
\end{eqnarray}
where   $H_{host}$   describes  carrier   propagation   in  the   host
disordered semiconductor,  
here approximated  by a  single band  Hamiltonian with
semicircular  density of  states,  the second  term  describes the
exchange coupling of  the carriers to an array of  Mn moments $S_i$ at
positions $R_i$, and the third term is the scalar part of the
carrier-Mn potential. \cite{Hwang:2002_} At zero temperature, when all 
Mn moments $S_i$ are aligned,  the carriers with spin parallel to $S_i$
feel  a   potential  $-J+W$  on   each  magnetic  impurity   site  and
anti-parallel carriers feel a potential $J+W$.  Within this model, the
disorder  is treated  in the  dynamical mean  field  approximation, or
equivalently  in  these   systems  the  dynamical  coherent  potential
approximation.\cite{Takahashi:2003_cond-mat/0306588} The  real part of
the conductivity is given by
\begin{eqnarray}
\sigma(\Omega,T)&=&\int \frac{d^3p}{(2\pi)^3} \left ( 
\frac{p\cos\theta}{m} \right )^2
\int \frac{d\omega }{\pi }\frac{\left[ f(\omega )-f(\omega +\Omega
)\right] }{\Omega }\nonumber \\
&\times& {\rm Im} G(p ,\omega ) {\rm Im} G(p ,\omega +\Omega).
\label{sigxx}
\end{eqnarray}

Typical  results for the  conductivity obtained  within this  model are
shown in Fig. 3 of Ref. ~\onlinecite{Hwang:2002_}.  Although the curves are  similar to the
ones  obtained using  the  effective Hamiltonian  multi-band models,  the
origin of the peak at  intermediate frequencies is quite different. In
the lattice model the peak corresponds to transitions between
the main band and the impurity band that forms when
the carrier-Mn coupling  becomes comparable or stronger
to the main band width  and spin-splitting.   
The  high  value  of  the  exchange
coupling required for this physics to apply is, however,  
not consistent with experimental $J_{pd}$ values
inferred from the visible-range MCD or magnetotransport
measurements.\cite{Ando:1998_,Ohno:1999_a}

Monte Carlo  simultations in closely related lattice  models exhibit a
similar   conductivity  behavior arising  again   from   the   presence
of the impurity band.\cite{Alvarez:2003_} 
These theories  point out the qualitative
features  present in  the  theoretical  models as  a  function of  the
material parameters which, although  fixed in most DMS materials, may
be  tuned by  chemical engineering. It  would  be of
interest to  extend the calculation to a  multi-band Hamiltonian model
which  may lead  to a  more complex description of  the intermediate
frequency regime. In addition,  this
extension  would allow  for a  calculation of  magneto-optical effects
within this models which are not available at present.

\subsection{Magneto-optical effects}
\label{MCD}
\begin{figure}
\includegraphics[width=3.4in]{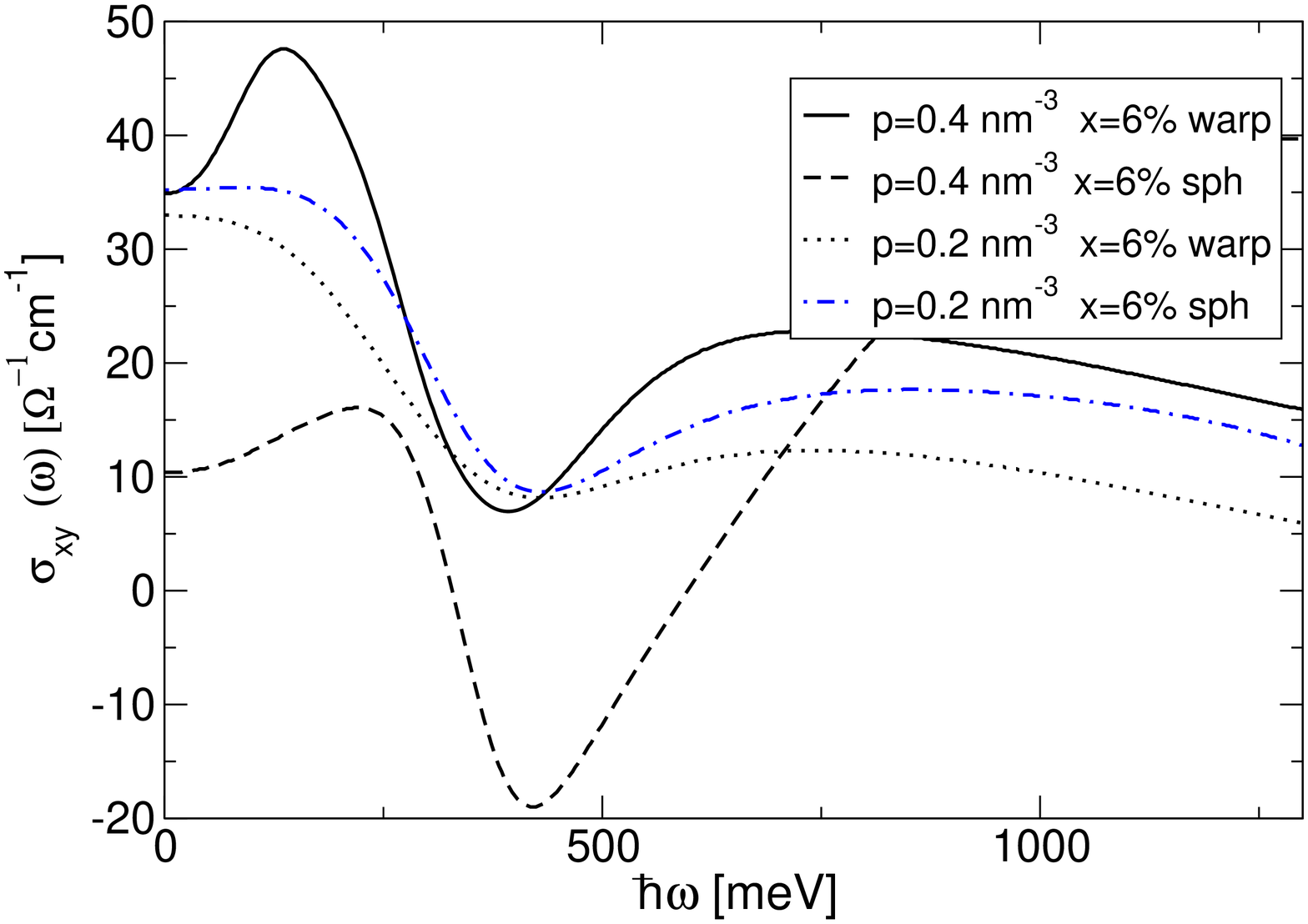}
\caption{Anomalous ac-Hall conductivity $\sigma_{xy}(\omega)$
for $x=6\%$ Mn concentration and $p=0.4$ and $0.2 {\rm nm}^{-3}$,
for spherical and non-spherical (band-warping) models. After Ref. 
57}
\label{sigAH_x6_p4}
\end{figure}

Magneto-optical  effects, such  as magnetic  circular  dichroism (MCD),
Kerr effect, and Faraday effect, give further insight in ferromagnetic
materials  and add  an aditional  insight for   the modeling  of the
electronic structure.   Absorption and reflection  measurements in the
visible range  have been used to  establish phenomenological estimates
for the p-d  and s-d exchange coupling constants  in DMS materials and
are found to be in agreement with magnetic susceptibility measurements
obtained  from  magneto-transport  measurements  in  the  paramagnetic
regime.   \cite{Dietl:2001_a,Ando:1998_,Szczytko:1999_a,Beschoten:1999_}
Measurements  of magneto-optical  coefficients on  band  energy scales
provide  very  detailed  information  about the  influence  of  broken
time-reversal symmetry on itinerant electron quasiparticle states. For
DMS materials  the corresponding energy scale for  the heavily p-doped
(III,Mn)V  ferromagnets  is in  the  infrared.   Within the  effective
Hamiltonian      model,       the      anomalous      Hall      effect
theory\cite{Jungwirth:2003_b} which has proven succesful when comparing
closely with experimental results, can be easily extended to 
finite frequency range:\cite{Sinova:2003_}
\begin{widetext}
\begin{eqnarray}
{\rm Re}[\sigma_{xy}(\omega)]&=&-\frac{e^2\hbar}{m^2 V }
\frac{d\vec{k}}{(2\pi)^3}
\sum_{\vec{k}n\ne n'}
(f_{n',\vec{k}}-f_{n,\vec{k}}) 
\nonumber\\ &\times& 
\frac{{\rm Im}[\langle n'
\vec{k}|\hat{p}_x|n\vec{k}\rangle\langle n\vec{k}| \hat{p}_y|n'\vec{k}
\rangle]
(\Gamma_{n,n'}^2+\omega(E_{n\vec{k}}-E_{n,\vec{k}'})
-(E_{n\vec{k}}-E_{n,\vec{k}'})^2)}
{((\omega-E_{n\vec{k}}+E_{n'\vec{k}})^2+(\hbar\Gamma_{n,n'})^2)
((E_{n\vec{k}}-E_{n'\vec{k}})^2+(\hbar\Gamma_{n,n'})^2)},
\label{ac_sig_AHE_dis}
\end{eqnarray}
\end{widetext}
where   $\Gamma_{n,n'}\equiv   (\Gamma_{n}+   \Gamma_{n'})/2   $   and
$\Gamma_n$ are the golden rule scattering rates averaged over band $n$
as shown in Sec.  \ref{boltzman} and \ref{absorb}.  Typical values for
this     anomlaous    ac-Hall     conductivity     are    shown     in
Fig.~\ref{sigAH_x6_p4}
    
The  Hall conductivity $\sigma\xy$ must be  non-zero  in order  to have  non-zero
magneto-optical effects, such as the  Faraday and Kerr effect, but the
full conductivity  tensor must be known  in order to connect the calculations with the
measurable  quantities.  The calculations and measurements mentioned here 
are for the Faraday geometry (magnetic field parallel to light propagation axis) 
for transmission and the polar Kerr geometry (near normal incidence with the magnetic field 
along the light propagation direction) for reflection.
The complex Faraday angle $\widetilde\theta\f$ includes both real and imaginary terms.  
$\widetilde\theta\f$ is defined as
\begin{equation}
\tan\widetilde\theta\f\equiv
\frac{t\xy}{t\xx}=\frac{t_+-t_-}{i(t_++t_-)}=\theta\f+i\eta\f,
\end{equation}
where $t\xx$ and $t\xy$ are the complex transmission
amplitudes for linearly polarized light and $t_\pm$ are 
the total complex transmission amplitudes (with multiple scattering 
taken into account) for right and left circular polarized light.  In typical measurements,
$\widetilde\theta\f$  is small and therefore the small
angle approximation $\tan\widetilde\theta\f\approx \widetilde\theta\f$ can be used.  
The real term (Re$[\widetilde\theta\f]$ or $\theta\f$) corresponds to a simple geometric 
rotation of the polarization vector about the direction of propagation.  
The imaginary term (Im$[\widetilde\theta\f]$ or $\eta\f$) relates directly to the 
ellipticity of the polarization.  If the sample is
axially symmetric along the magnetic field $B$, the transmittance tensor is diagonal when
represented in the circular polarization basis.  Therefore, changes in the incident
polarization only depend on: (1) the  relative difference in the phase of left
versus right circularly polarized light due to 
Re$[\widetilde\theta\f]$, 
which leads to a rotation (circular birefringence or Faraday
rotation, FR) in the linearly polarized incident light; and (2) the relative
difference in the transmission of left versus right circularly polarized 
light due to Im$[\widetilde\theta\f]$,
which introduces ellipticity (circular dichroism, CD) to the linearly polarized incident light.

In the thin film approximation, the relationship
between $\theta\f$ and and the conductivity is given by:
\begin{eqnarray}
\tan \theta\f\equiv\frac{t\xy}{t\xx}
&=&\biggl(1+\frac{1}{Z\sigma\xx}\biggr)\frac{\sigma\xy}{\sigma\xx}
\label{eq;hallfar}
\end{eqnarray}
where $Z= Z_0 d/(n+1)$, $Z_0$ is the impedance of free space, $n$ is the
substrate index of refraction, and $d$ is the film thickness.  In bulk, the angle of  rotation per
unit length traversed is
\begin{equation}
\theta_F(\omega)=\frac{4\pi}{(1+n)c}{\rm Re}[\sigma_{xy}],
\end{equation}
where $c$ is the speed of light  and $n$ is the index of refraction of
the  substrate, in  this  case GaAs  with  $n=\sqrt{10.9}$.  

CD or MCD (magnetic circular dichroism) is related to the 
difference between the optical absorption of right and left circularly
polarized light,  and is given by
\begin{equation}
MCD=\frac{\alpha^+-\alpha^-}{\alpha^++\alpha^-}=\frac{{\rm
Im}[\sigma_{xy}(\omega)]} {{\rm Re}[\sigma_{xx}(\omega)]}
\propto {\rm Im}[\widetilde\theta\f].
\end{equation}
where $\alpha_\pm$ are the absorbances for left and right circularly 
polarized light.
At visible range frequencies multiscattering reflections must be taken
into account since in that  regime the wavelength is comparable to the
typical epilayer thickness.

Whereas the complex Faraday angle describes the polarization of a transmitted beam, the complex Kerr angle $\widetilde\theta\k$ is applied in the same way for reflected radiation, with
\begin{equation}
\tan\widetilde\theta\k\equiv\frac{r\xy}{r\xx}
=\frac{r_+-r_-}{i(r_++r_-)}=\theta\k+i\eta\k,
\end{equation}
where $r\xx$ and $r\xy$ are the complex reflection amplitudes for linearly polarized light and $r_\pm$ are the total complex reflection amplitudes (with multiple scattering 
taken into account) for right and left circular polarized light. 

\begin{figure}
\includegraphics[width=3.2in]{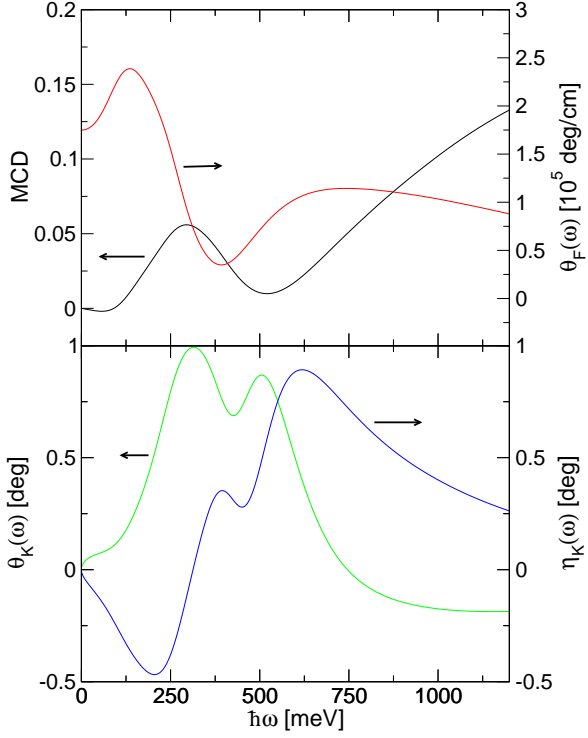}
\caption{Faraday and Kerr effects 
for $x=6\%$ Mn concentration and $p=0.4$ ${\rm nm}^{-3}$. 
After Ref. 57}
\label{mo_eff_x6_p4}
\end{figure}
Fig.  \ref{mo_eff_x6_p4}  shows  the different  magneto-optic  effects
predicted   for  a   concentration  of   $x=6\%$  and   $p=0.4$  ${\rm
nm}^{-3}$. The Faraday rotation in  this case is larger than the giant
Faraday rotation  observed in the paramagnetic  (II,Mn)VI's at optical
frequencies  \cite{Sugano:2000_} and should  be readily  observable in
the highly metallic samples.

\begin{figure}
\includegraphics[width=3.5in]{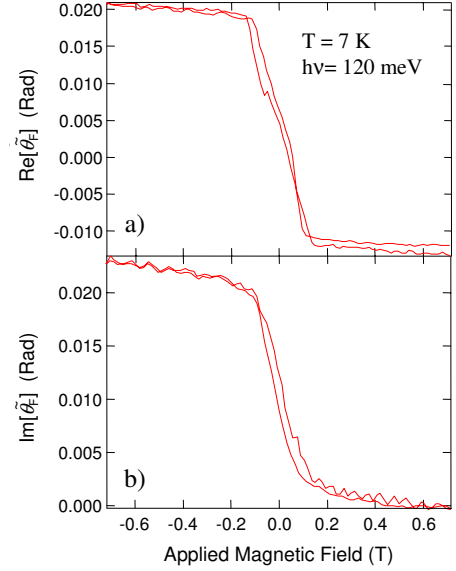}
\caption{Re$[\tilde\theta\f]$ a) and Im$[\tilde\theta\f]$ 
in the MIR range b) as a function of applied magnetic field for 
GaAs(6~\% Mn) random alloy epilayer at 7~K and probe radiation 
energy 118 meV, after Ref. 122}
\label{faraday}
\end{figure}

\begin{figure}
\includegraphics[width=3.3in]{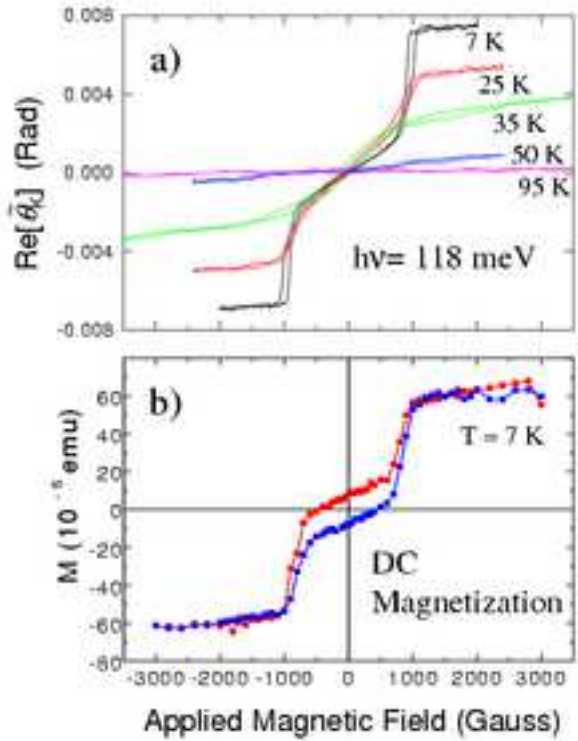}
\caption{Re$[\tilde\theta\k]$ in the MIR (118 meV) and dc magnetization (courtesy of S. Wang) for GaAs(1~\% Mn) random alloy epilayer.  After Ref. 122}
\label{kerr}
\end{figure}

Though magneto-optical measurements on II-VI DMS materials have 
been extensive, see for example Ref. ~\onlinecite{Ramdas:1998_},
measurements on III-V DMS have been more limited, with almost no Faraday/Kerr measurements
reported on these compounds in the mid-infrared (MIR, 100-500 meV) range. 
The MIR range provides a critical test of the predictions in Ref.~~\onlinecite{Sinova:2002_}.  
Transmission and reflection MIR magneto-optic meaurements on Ga$_{1-x}$Mn$_x$As random alloy films 
in Figs. \ref{faraday} and \ref{kerr}, respectively have been recently measured 
in this MIR range and support the presence of such strong magneto-optic signal in the MIR regime.\cite{Cerne:2004_}

Figure~\ref{faraday} plots the real a) and imagnary b) parts of the MIR $[\widetilde\theta\f]$ 
probed at a photon energy of 118 ~meV and a sample temperature of 7~K.  
The hysteresis is weak and similar to out-of-plane dc magnetization measurements performed on this sample  
(a 6~\% Ga$_{1-x}$Mn$_x$As random alloy film). 
Both the real and imaginary parts of $[\widetilde\theta\f]$  appear to be 
proportional to the film magnetization, as expected.  
The Re$[\tilde\theta\f]$ signal at magnetization saturation calculated by 
Sinova {\it et al.} \cite{Sinova:2002_} are on the order of 0.04 rad for a film with a 
similar thickness and Mn concentration. However such calculation are done in optimally annealed samples and
further experimental measurements in this more metallic regime will be helpful to establish the validity
of the model.

In Fig.~\ref{kerr}, a two-step hysteresis loop is observed both the MIR 
(118 meV) Re$[\widetilde\theta\k]$ as well as the dc magnetization for 
a 1~\% Ga$_{1-x}$Mn$_x$As random alloy film. This behavior is similar to that 
observed in MCD at a photon energy of 2.83~eV by Ref.~~\onlinecite{Ando:1998_}.
The authors in Ref.~~\onlinecite{Ando:1998_} claim that such behavior cannot be attributed 
to GaMn clusters, and therefore is a signature of the Ga$_{1-x}$Mn$_x$As alloy.  

The MIR magneto-optical measurement technique used in the above measurements is described in 
detail in Ref. ~\onlinecite{Cerne:2003_}, and results using this technique 
on non-magnetic systems are presented in Refs. ~\onlinecite{Cerne:2000_a} 
and ~\onlinecite{Cerne:2000_b}.  The results shown in Fig. ~\ref{faraday} and ~\ref{kerr} 
serve to demonstrate the experminetal feasibility of such studies, but 
further work is required to interpret these results and to compare them with theoretical predictions.

\section{Summary}

This review  has focused on highliting the rapid  developments in  
material research  of  metallic ferromagnetic (III,Mn)V semiconductors over  
the past few years. Although the successes have been extensive 
in the understanding and development of  the  bulk
properties  of the (Ga,Mn)As and (In,Mn)As systems within  the metallic  regime, many challenges remain
both in other (III,Mn)V based semiconductors and lower dimensionalities such
as digitally doped heterostructures. Both transport and magneto-optical effects
will remain a very fruitfuil ground to make progress in these materials. At the time of 
writing, there is still a lot of work being done trying to sort out the more complicated
physics of (Ga,Mn)N and several reports of high $T_c$ measurements in the digitally doped
materials. We simply keep watching in fascination the rapid developments and challenges
posed by new experiments which will undoubtedly generate interesting physics and new
understandings of these fascinating materials.

\section*{Acknowledgements}
The many collaborators and colleagues with whom we have had the privilage to work with
and learn from are too numerous to name here and we are greatful for their contribution
to our knowledge of the subject. However, the authors would like to 
acknowledge insightful discussions during the course of this review with 
W. Atkinson, D. Basov, K. Burch, T. Dietl, 
J. Furdyna, H. Luo, J. Macek, A. H. MacDonald, B. McCombe, and C. Timm.
This work was supported in part by the Welch Foundation,
DOE under grant DE-FG03-02ER45958, and
the Grant Agency of the Czech Republic
under grant 202/02/0912.


\begin{thebibliography}{125}
\bibitem{noteonAHE}
\bibinfo{note}{Sometimes it is called extraordinary or spin Hall effect,
  although these names are now in disuse and the later one is now used in a
  different context within paramagnetic materials.}
\end{thebibliography}
\end{document}